\documentclass[twocolumn,prd,aps,groupedaddress,showpacs,nofootinbib,preprintnumbers]{revtex4}

\usepackage{amsmath}
\usepackage{amsfonts}
\usepackage{amssymb}
\usepackage{url}
\usepackage{graphicx}
\usepackage{bm}
\usepackage{psfrag}

\def\be{\begin{equation}}
\def\ee{\end{equation}}
\def\ba{\begin{eqnarray}}
\def\ea{\end{eqnarray}}
\def\bs{\begin{subequations}}
\def\es{\end{subequations}}
\def\cS{{\cal S}}

\def\B{\Box}
\def\a{\alpha}
\def\b{\beta}

\def\s{\sigma}

\def\p{\partial}
\def\tphi{\tilde\phi}
\def\cO{{\cal O}}

\newcommand{\Eq}[1]{(\ref{#1})}
\def\rmi{i}
\def\rme{e}
\def\rmd{d}
\def\Im{{\rm Im}}

\begin{document}
\title{Route to nonlocal cosmology}
\author{Gianluca Calcagni}
\email{g.calcagni@sussex.ac.uk}
\affiliation{Department of Physics and Astronomy, University of Sussex, Brighton BN1 9QH, United Kingdom}
\author{Michele Montobbio}
\email{montobb@science.unitn.it}
\affiliation{Dipartimento di Fisica and INFN Gruppo Collegato di Trento, Universit\`a di Trento, via Sommarive 14, 38100 Povo (Trento), Italia}
\author{Giuseppe Nardelli}
\email{nardelli@dmf.unicatt.it}
\affiliation{Dipartimento di Matematica e Fisica, Universit\`a Cattolica,
via Musei 41, 25121 Brescia, Italia\\
and INFN Gruppo
Collegato di Trento, Universit\`a di Trento, 38100 Povo (Trento), Italia}
\date{May 21, 2007}

\begin{abstract}
An analytic approach to phenomenological models inspired by cubic string field theory is introduced and applied to some examples. We study a class of actions for a minimally coupled, homogeneous scalar field whose energy density contains infinitely many time derivatives. These nonlocal systems are systematically localized and an algorithm to find cosmological solutions of the dynamical equations is provided. Our formalism is able to define the nonlocal field in regions of the parameter space which are inaccessible by standard methods. Also, problems related to nonlocality are reinterpreted under a novel perspective and naturally overcome. We consider phenomenological models living on a Friedmann--Robertson--Walker background with power-law scale factor, both in four dimensions and on a high-energy braneworld. The quest for solutions unravels general features of nonlocal dynamics indicating several future directions of investigation.
\end{abstract}


\preprint{PHYSICAL REVIEW D \textbf{76}, 126001 (2007) \hspace{9cm} arXiv:0705.3043 [hep-th]}
\maketitle


\section{Introduction}

The interest in nonperturbative formulations of string theory has been increasing for two complementary reasons. From a theoretical point of view, the dynamics of strongly interacting strings and its consequences in the articulation of a fundamental, predictive theory are far from being fully explored. In parallel, cosmology is a natural field for testing string theory, since observations at very large scales (e.g., of the cosmic microwave background) probe energies close to the Planck scale. As soon as new developments are achieved on the theoretical side, phenomenological details are gathered on their cosmological aspects. Ultimately, we hope that this process is two way and cosmology will be able to indicate promising research directions to the string community.

One framework falling into the above discussion is string field theory, particularly in its cubic version (CSFT); see \cite{sen04,ohm01,ABGKM} for some reviews. The effective action of the open CSFT particle modes is \emph{nonlocal}, in the sense that the fields are derived infinitely many times by a set of d'Alembertian operators of the form $\rme^{r_*\B}=1+r_*\B+r_*^2\B^2/2+\dots$, where $r_*$ is a number fixed by the theory. The simplest (and best studied) example of such effective actions is obtained when all fields are switched off except for the scalar tachyon. In the Minkowski case, one can find that approximate (although very accurate) solutions satisfy a diffusion equation with respect to an auxiliary evolution variable $r$, eventually set equal to the physical value $r_*$ \cite{FGN,roll}.

One can try to apply this procedure also in the presence of gravity. There are a number of reasons why to consider the cubic SFT tachyon also on a cosmological [Friedmann--Robertson--Walker (FRW)] background. First, nonlocal theories are still poorly understood and it would be instructive to embed them in an expanding universe and compare the resulting cosmology with the standard big bang/inflationary model. Second, both the bosonic and supersymmetric Minkowski solutions found in \cite{FGN,roll} have a spike at the origin of time.\footnote{Unless one promotes the regulator in the solutions as a physical parameter; in this case, the scalar field is a regular function of time.} One might ask whether a Hubble friction term would modify this behaviour in the FRW extension of such solutions.

Third, by now a cosmological tachyon has been extensively studied in its Dirac--Born--Infeld (DBI) formulation. The first version of the DBI effective action \cite{gib02} (see \cite{CGT,PhD} and references therein for an overview of the literature) is valid when higher-than-first derivatives of the field are small compared to its speed \cite{gar00,bers0,klu00,GHY}. The simplest versions of tachyon inflation do not reheat and generate large density perturbations, unless the parameters of the model are fine-tuned \cite{CGJP,FrKS,KL,SCQ,PCZZ,BBS,BSS,PS,rae04}. Although these problems can be bypassed \cite{GST}, it is unlikely that even a viable tachyon would lead to predictions qualitatively different from those of standard inflation \cite{PhD}. Moreover, DBI actions tend to develop caustics on an inhomogeneous background, which break down the DBI approximation (higher derivatives diverge on caustics) \cite{FKS}. It was also shown that the field is unsatisfactory as model of quintessence \cite{CL}.\footnote{In general, the string tachyon is regarded as a late-time candidate for dark energy only on phenomenological grounds, because its evolution scale is of order of the string scale. Tachyon dynamics can be associated more naturally to the very early universe.} In a more recent and promising formulation of DBI scenarios, higher-order derivatives are taken into account and new features, such as large non-Gaussianities, can emerge \cite{dbi01,dbi02,dbi12,dbi13,dbi14,dbi03,dbi04,dbi05,dbi06,dbi07,dbi08,dbi09,dbi10,dbi11}.

There is evidence that the cubic SFT tachyon is not plagued by the above fine-tunings and may give interesting phenomenology \cite{are04,AJ,cutac,AK,AV,BBC,kos07,AKV,AV2,lid07,BC}. However, there are several conceptual issues \cite{cutac} which have not yet been addressed. It is our purpose to verify these issues and comment on past results. 
We start this research program from the study of phenomenological models, presenting the detailed CSFT tachyon case in a forthcoming publication \cite{cuta3}.

It will be convenient to inspect the effective cosmology on Randall--Sundrum (RS) and Gauss--Bonnet (GB) four-dimensional braneworlds, in the approximation of high energy and no brane-bulk energy exchange (see \cite{PhD,cutac} for an overview of the braneworld formalism used here\footnote{This approximation at the level of the background equations was invoked also to simplify the analysis of large-scale inflationary perturbations. The legitimacy of this attitude has been recently validated by the results of the full computation of Ref.~\cite{KMRWH}.}). The local equations are easily generalized to these cases \cite{PhD}, which are interesting in their own because the superstring tachyon mode naturally lives on a non-BPS brane \cite{ohm01}.

It is important to stress that the methodology of this paper and its companions \cite{roll,cuta3}, which has tight connections with the $1+1$ Hamiltonian formalism developed in \cite{LV,GKL,beri0,CHY,GKR}, is of particular relevance when the scalar field is interpreted as the inflaton. There, the field is quantized and perturbed to compute its $n$-point correlation functions (primordial $(n-1)$-spectra), an aspect which will be not dealt with here. Without localizing the derivative operators, it is difficult to define a set of fields, conjugate momenta and commutation relations in the usual Ostrogradski way. To the best of our knowledge, there are no other known ways to achieve a sensible quantization while retaining the genuine nonlocal nature of the problem.

Another benefit of our proposal is that the nonlocal Cauchy problem \cite{MZ} becomes transparent and reduces to the determination of a finite set of parameters of the candidate solution. The reader is referred to Eqs.~\Eq{exeo1}--\Eq{exeo3} for a braneworld example.

This is the outline of the paper: Section \ref{setup} contains some review material on cubic SFT and nonlocality on curved (in particular, FRW) backgrounds. In Sec.~\ref{exloc} exact solutions of the local system are reviewed. Section \ref{covan} describes an attempt to find nontrivial cosmological solutions via a covariant \emph{Ansatz}, while in Sec.~\ref{series} it is shown that nonlocal solutions as infinite power series cannot be constructed from exact local cosmological solutions. Section \ref{strat} explains the strategy to find approximated nonlocal solutions with the local solutions as initial conditions in an auxiliary evolution variable. These steps are applied concretely in Sec.~\ref{1tbw} for braneworld cosmologies with constant slow-roll parameters; there, we confront the truncated power series of Sec.~\ref{series} with the ``localized'' function thus found. The same procedure is applied in Sec.~\ref{1t4d} for 4D power-law cosmology. A discussion of the results, comparison with literature, and open problems are in Sec.~\ref{disc}. The Appendix contains technical material.


\section{Setup}\label{setup}

The bosonic cubic SFT action is of Chern--Simons type \cite{wi86a},
\be\label{SFT}
\cS=-\frac{1}{g_o^2}\int \left(\frac{1}{2\alpha'} \Phi* Q_B\Phi+\frac13\Phi*\Phi *\Phi\right),
\ee
where $g_o$ is the open string coupling constant (with $[g_o^2]=E^{6-D}$ in $D$ dimensions), $\int$ is the path integral over matter and ghost fields, $Q_B$ is the Becchi--Rouet--Stora--Tyutin operator, * is a noncommutative product, and the string field $\Phi$ is a linear superposition of states whose coefficients correspond to the particle fields of the string spectrum.

At the lowest truncation level \cite{MSZ}, all the excitations in $\Phi$ are neglected except for the tachyon field, labelled $\phi(x)$ and depending on the center-of-mass coordinate $x$ of the string. The Fock-space expansion of the string field is truncated so that $\Phi \cong |\Phi\rangle=\phi(x)|\!\downarrow\rangle$, where the first step indicates the state-vertex operator isomorphism and $|\!\downarrow\rangle$ is the ghost vacuum with ghost number $-1/2$.
At level $(0,0)$ the action becomes, in $D=26$ dimensions and with metric signature $({-}{+}{\dots}{+})$ \cite{KS1,KS2},
\ba
\bar{\cS}_\phi &=&\frac{1}{g_o^2}\int \rmd^D x \left[\frac{1}{2\alpha'}\phi(\alpha'\p_\mu\p^\mu+1)\phi\right.\nonumber\\
&&\quad\left.
-\frac{\lambda}{3}\left(\lambda^{\alpha'\p_\mu\p^\mu/3}\phi\right)^3-\Lambda\right],\label{tactmin}
\ea
where $\lambda=3^{9/2}/2^6\approx 2.19$, $\alpha'$ is the Regge slope, and Greek indices run from 0 to $D-1$ and are raised and lowered via the Minkowski metric $\eta_{\mu\nu}$. The tachyon field is a real scalar with dimension $[\phi]=E^2$. The constant $\Lambda$ does not contribute to the scalar equation of motion but it does determine the energy level of the field. In particular, it corresponds to the $D$-brane tension which sets the height of the tachyon potential at the (closed-string vacuum) minimum to zero. This happens when $\Lambda=(6\lambda^2)^{-1}$, which is around $68\%$ of the brane tension; this value is lifted up when taking into account higher-level fields in the truncation scheme.

On a curved background, we define the operator (in $\alpha'$ units)
\be\label{lb}
\lambda^{\B/3}= \rme^{r_*\B} \equiv\sum_{\ell=0}^{+\infty}\frac{(\ln\lambda)^\ell}{3^\ell \ell!} \B^\ell=\sum_{\ell=0}^{+\infty}c_\ell \B^\ell\,,
\ee
where
\be\label{dal}
\Box \equiv \frac{1}{\sqrt{-g}}\,\p^\mu (\sqrt{-g}\,\p_\mu),
\ee
and
\be
r_*\equiv \frac{\ln\lambda}{3}=c_1=\ln 3^{3/2}-\ln 4\approx 0.2616.
\ee
Defining the ``dressed'' scalar field
\be\label{dres}
\tphi\equiv \lambda^{\B/3}\phi=\rme^{r_*\B}\phi,
\ee
and coupling the tachyon to a general metric $g_{\mu\nu}$, the total action is
\be\label{tact}
\cS=\int \rmd^D x \sqrt{-g}\left[\frac{R}{2\kappa^2}+\frac12\,\phi(\B-m^2)\phi-U(\tphi)-\Lambda\right],
\ee
where $g$ is the determinant of the metric, $R$ is the spacetime Ricci scalar, $\kappa$ is the effective gravitational coupling, $m^2$ is the squared mass of the scalar field (negative for the tachyon), $U$ is a nonlocal term that generalizes the cubic potential of Eq.~\Eq{tactmin}, and we have absorbed the open string coupling into $\phi$. For simplicity we have written the gravitational part as the Einstein--Hilbert action, although later on we shall discuss other possibilities.

We specialize to a four-dimensional target where the other dimensions are assumed to be compactified, stabilized, and integrated out. On the flat FRW metric $\rmd s^2=-\rmd t^2+a^2(t)\, \rmd x_i \rmd x^i$, for a spatially
homogeneous field $\phi\equiv\phi(t)$ the d'Alembertian operator reduces to
\be
\B=-(\rmd_t^2+3H\rmd_t)\,,
\ee
where $t$ is synchronous time, $a(t)$ is the scale factor, $H\equiv \dot a/a$ is the Hubble parameter, and $\rmd_t=\dot{}\,$ is the derivative with respect to $t$. All the formul\ae\ below reduce to the Minkowski case for $H=0$.

Following the notation of \cite{cutac} (to which we refer for the detailed derivation of the dynamical equations), the equation of motion for the nonlocal scalar field  is
\be\label{teom}
\B\phi=m^2\phi+U'\,,
\ee
where
\be
U'\equiv\rme^{r_*\B}\tilde U'\equiv\rme^{r_*\B}\frac{\p U}{\p\tphi}
\ee
is constructed from a nonlocal potential term $U(\tphi)$ which does not contain derivatives of $\tphi$. One can also recast Eq.~\Eq{teom} in terms of $\tphi$,
\be\label{teom2}
(\B-m^2) \rme^{-2r_*\B}\tphi=\tilde U'.
\ee
In the string cases the potential is monomial, $U(\tphi)=\sigma(r_*)\,\tphi^n$, where $\sigma(r_*)$ is a coupling constant ($n=3$ for the bosonic open string, $n=4$ for the approximate supersymmetric string). The total  potential is
\be\label{pot}
\tilde V(\phi,\tphi) \equiv \frac{1}{2}m^2\phi^2+U(\tphi)+\Lambda\,.
\ee
On a FRW background there is an extra equation relating the gravitational sector to the scalar one:
\be\label{FRW}
f_\textsc{frw}(t)\equiv H^2-\frac{\kappa^2}{3}\,\left[\frac{\dot{\phi}^2}{2}(1-\cO_2)+\tilde V -\cO_1\right]=0,
\ee
where
\ba
\cO_1 &=& r_*\int_0^{1} \rmd s\, (\rme^{s r_*\B}\tilde U')(\B \rme^{-s r_*\B}\tphi)\nonumber\\
      &=& r_*\int_0^{1} \rmd s\, [(\B-m^2)\rme^{(s-2) r_*\B}\tphi](\B \rme^{-s r_*\B}\tphi),\label{O1}\\
\cO_2 &=& \frac{2r_*}{\dot{\phi}^2}\int_0^{1} \rmd s\, \p_t(\rme^{sr_*\B}\tilde U')\,\p_t(\rme^{-sr_*\B}\tphi)\nonumber\\
      &=& \frac{2r_*}{\dot{\phi}^2}\int_0^{1} \rmd s\, \p_t[(\B-m^2)\rme^{(s-2) r_*\B}\tphi]\,\p_t(\rme^{-sr_*\B}\tphi) .\nonumber\\\label{O2}
\ea
In the local case ($r_*=0$, $\lambda=1$), $\cO_i=0$.

It is straightforward to generalize Eq.~\Eq{FRW} to scenarios where the visible universe lives on a (3+1)-brane embedded in a higher-dimensional spacetime. If one neglects energy exchange between brane and bulk modes, the only modification to the equations of motion is in the Friedmann equation \Eq{FRW}, which becomes, at high energies,
\be\label{FRWth}
H^{2-\theta}-\beta_\theta^2 \rho= 0\,.
\ee
The constants $\theta$ and $\beta_\theta$ are determined by the brane setup. In particular, for a Randall--Sundrum brane $\theta=1$ \cite{BDL,CGKT,CGS,BDEL,FTW}, while for a brane in Gauss--Bonnet gravity $\theta=-1$ \cite{CD,dav03,GW}. At low energies, one recovers general relativity. We shall comment on the approximations implied by Eq.~\Eq{FRWth} in the final section.


\section{Exact local solutions}\label{exloc}

Before considering the nonlocal system, we recall some exact solutions of the standard local equations of motion ($r_*=0$, $\tphi=\phi$) with $m^2=0$ (mass terms can reappear in $U$). We start from a configuration
\ba
\phi &=&\phi_0  t^p,\label{fp}\\
H    &=& H_0 t^q, \label{hq}
\ea
where $p$ and $q$ are real numbers and $\phi_0$, $H_0$ are constants. \\ This \emph{Ansatz} parametrizes two widely studied classes of exact cosmological solutions of the local equations, both in general relativity and other theories of gravity such as the braneworld.

The first class has a power-law scale factor ($q=-1$, $a=t^{H_0}$), corresponding, in four dimensions ($\theta=0$) and modulo integration constants, to \cite{AW,LM2}
\ba
\phi &=& \sqrt{\frac{2H_0}{\kappa^2}}\,\ln t\,,\\
U &=& \frac{(3H_0-1)H_0}{\kappa^2} \exp \left(-\sqrt{\frac{2}{H_0}}\,\kappa\phi\right)\,.
\ea
In the Randall--Sundrum ($\theta=1$) and Gauss--Bonnet ($\theta=-1$) braneworlds, 
\ba
\phi &=& \phi_0\, t^{\theta/2}=\frac2\theta \sqrt{\frac{2-\theta}{3\beta_\theta^2}} H_0^{(1-\theta)/2}\,t^{\theta/2}\,,\\
U &=& \left(1-\frac{2-\theta}{6H_0}\right)\frac{H_0^{2-\theta}}{\beta_\theta^2} \left(\frac{\phi}{\phi_0}\right)^{2(1-2/\theta)}\equiv \sigma\phi^{2(1-2/\theta)}\,,\nonumber\\\label{beo}
\ea
which are $\phi \propto t^{1/2}$, $U\propto \phi^{-2}$ and $\phi \propto t^{-1/2}$, $U\propto \phi^{6}$, respectively (see \cite{cal3} for details).

Another exact solution reads
\be \label{aexp}
a(t)= \exp (p\, t^n)\,,\quad H=pn\,t^{n-1}\,,\quad \text{sgn}(p)=\text{sgn}(n)\,,
\ee
which gives
\ba
\phi &=& \phi_0\, t^{[n(1-\theta)+\theta]/2}\,,\\
U &=& A
\phi^{\frac{2(n-1)(2-\theta)-2n}{n(1-\theta)+\theta}}+B\phi^{\frac{2(n-1)(2-\theta)}{n(1-\theta)+\theta}}\label{bwpote}\,,
\ea
for some $A=A(n,p)$ and $B=B(n,p)$ \cite{cal3,bar90,BS}. There is also a Gauss--Bonnet solution $\phi \propto \ln t$ with exponential potential when $n=1/2$ \cite{cal3}, which we shall not consider here. Finally, the case $n=2$ (linear Hubble parameter) may be of particular interest because in four dimensions it corresponds to a quadratic potential: 
$\phi = \phi_0 t^{1-\theta/2}$, $U =A \phi^{2\theta/(\theta-2)}+ B\phi^2$.

We note that it is possible to formally get a potential of type $A\phi^2+B\phi^4$, $A\neq0\neq B$, only in two situations. The first is for $\theta=0$ in the limit $n\to \infty$ in Eq.~\Eq{bwpote}, where the coefficients of the potential as well as $a$ and $H$ are ill-defined. The second is an effective scenario (with no connection to any known modified gravity model) with $\theta=4$ and $n=2$, which has $\phi=\phi_0/t$ and a Hubble parameter linear in $t$. The effective potential is then 
\be
U =3p(\phi^2-2\beta^2_4p^2\phi^4)\,,
\ee
which has an extra $-$ sign relative to the string potential if the universe expands and $\phi$ is real ($p>0$, $\beta_4^2>0$). This has the shape of a $(p=3)$-adic potential.


\section{Nonlocal solutions: Covariant \emph{Ansatz}} \label{covan}

The simplest way to solve the equations of motion is by requiring that
\be\label{ans1}
\B\phi=U_0+A\phi,
\ee
where $U_0$ and $A$ are constants. This \emph{Ansatz}, first introduced in \cite{BMS} for nonlocal gravitational degrees of freedom, has the advantage of being covariant, and hence independent from any specific background. It does not require knowing the Hubble parameter $H(t)$ to compute any nonlocal function like the $\cO_i$'s. The aim of this exercise is twofold. On one hand, it will show that there exist solutions of the nonlocal equations which are trivial, inasmuch as they behave just like local solutions of local models with rescaled parameters. On the other hand, by following Eq.~\Eq{ans1} we will be forced to abandon string theory and limit ourselves to a phenomenological model, letting the field mass be a free parameter and the nonlocal potential be unspecified. The simplifications given by Eq.~\Eq{ans1} are remarkable,\footnote{With a quadratic potential, the rescaling of the parameters can be used to heal the $\eta$ problem and obtain scale-invariant perturbation spectra also from local potentials which are usually too steep to support slow rolling. The effective parameters of the model are chosen to tune the slope of the potential while the physical ones maintain natural values \cite{lid07}.} but globally incompatible with string models. The key point is that the eigenvalue equation of the d'Alembertian operator is still the right path towards an analytic treatment of nonlocal systems, but under a different interpretation than that of Eq.~\Eq{ans1}. As we shall see, the eigenfunction of the $\B$ will not be the solution itself, but the kernel of an integral representation of the latter.

Comparing Eqs.~\Eq{teom} and \Eq{ans1} one has that $A = m^2$ and $U'=U_0$. Also, noticing that $\B^n\phi=(U_0+m^2\phi)(m^2)^{n-1}$, one gets
\ba
\tphi &=& U_0(1+\rme^{r_* m^2})+m^2 \rme^{r_* m^2}\phi\nonumber\\&=&U_0+\rme^{r_* m^2}\B\phi,\\
\B\tphi &=& m^2(\tphi-U_0)\,.
\ea
The last equation and Eq.~\Eq{teom2} imply that $\tilde U'=U'=B(U_0,m^2)$, $U=B(U_0,m^2)\phi+C$, where $C$ is an integration constant and
\be
B(U_0,m^2) \equiv -m^2 U_0(\rme^{-2r_* m^2}+1+\rme^{r_* m^2}).
\ee
Hence,
\be
\cO_1 = B(U_0,m^2) (\rme^{r_* m^2}-1)(U_0+m^2\phi),\qquad \cO_2 = 0,
\ee
so that the system reduces to a standard, local chaotic inflationary model with energy density
\ba
&&\rho=\frac{\dot{\phi}^2}{2}+\frac{1}{2}m^2\phi^2+D(U_0,m^2)\phi+\Lambda(U_0,m^2,C),\nonumber\\
&&D(U_0,m^2)       \equiv B(U_0,m^2)[1+(1-\rme^{r_* m^2})m^2],\nonumber\\
&&\Lambda(C,U_0,m^2) \equiv C+B(U_0,m^2) (1-\rme^{r_* m^2})U_0.
\ea
One can eventually choose $C$ so that $\Lambda=0$. This system has three free parameters $\{U_0,m^2,C\}$ as the standard phenomenological model. 

An example of a trivially local asymptotic solution is de Sitter evolution with $H=\sqrt{C}$ and $\phi\propto \rme^{-t}$. More generally, one can assume an exponential \emph{Ansatz} in de Sitter ($H=H_0$):
\be
\phi=\sum_{n=1}^{n_*} a_n \rme^{b_n t},
\ee
so that
\be
\tphi=\sum_{n=1}^{n_*} a_n \rme^{-r_* b_n(b_n+3H_0) t}\,,
\ee
and the scalar equation of motion reads
\ba
&&-\sum_{n=1}^{n_*} a_n [b_n(b_n+3H_0)+m^2]\rme^{b_n t}\nonumber\\
&&=\sigma \rme^{r_*\B}\left[\sum_{n=1}^{n_*} a_n \rme^{-r_* b_n(b_n+3H_0) t}\right]^k.
\ea
The left-hand side vanishes if
\be\label{bn}
|b_n|=\frac32 H_0\left[1\pm \sqrt{1-\left(\frac23\frac{m}{H_0}\right)^2}\right].
\ee
Then
\be
\sum_{n=1}^{n_*} a_n  =0,\nonumber
\ee
and one has local chaotic inflation ($U(\tphi)=$ const, the functions $\cO_i$ vanish). The Friedmann equation determines a set of algebraic relations between $a_n$ and $b_n$ coefficients, leaving freedom to choose $n_*-2$ coefficients $a_n$.


\section{Nonlocal solutions: Direct construction}\label{series}

As said above, in general one has to specify one independent \emph{Ansatz} for each degree of freedom, $\phi(t)$ and $H(t)$.
The procedure is inefficient and to find nontrivial solutions from this double prescription can be rather difficult. Here are a few examples.


\subsection{de Sitter case ($q=0$), $\phi=t^p$}

Although an exactly constant Hubble parameter requires the scalar field to be nondynamical, the dS solution is a good description of the standard inflationary regime.

In order to find $\tphi$, one has to act with multiple $\B$ operators on the scalar field. It can be shown that
\ba
\B^\ell\phi &=&(-1)^\ell\sum_{n=0}^{\ell}(3H_0)^{\ell-n}\left[\prod_{k=0}^{n+\ell-1}(p-k)\right]{\ell\choose n}
t^{p-n-\ell}\label{eq1}\nonumber\\
&=& \frac{\Gamma(\ell-p)\Psi(-\ell,1-2\ell+p;-3H_0t)}{\Gamma(-p)}(-1)^\ell\,t^{p-2\ell},\nonumber\\
\ea
where $\Psi$ is the confluent hypergeometric function of the second kind described in the Appendix. For fixed $t$ and $H_0$, $\Psi\to 1$ when $\ell\to 0$. The series defining $\tphi$ is not pointwise convergent, as can be checked by estimating the ratio
\ba
\frac{c_{\ell+1}\B^{\ell+1}\phi}{c_\ell\B^{\ell}\phi}&=&
c_1t^{-2}\frac{\Gamma(1+\ell-p)}{(\ell+1)\Gamma(\ell-p)}\nonumber\\
&&\times
\frac{\Psi(-1-\ell,1-2\ell-2+p;-3H_0t)}{\Psi(-\ell,1-2\ell+p;-3H_0t)}\nonumber\\
&\propto& \frac{(\ell-p)\Psi(-1-\ell,1-2\ell-2+p;-3H_0t)}{(\ell+1)\Psi(-\ell,1-2\ell+p;-3H_0t)}\nonumber\\
&\sim& \ell\nonumber
\ea
at large $\ell$, for finite $H_0$ and $t$. The last step can be checked numerically or conveniently worked out by taking the dominant term $n=\ell$ in Eq.~\Eq{eq1}. For large $\ell$ and fixed $p$, the product can be approximated as
\[
\prod_{k=0}^{2\ell-1}(p-k)\sim \Gamma(2\ell);
\]
then $\Gamma(2\ell+2)/[(1+\ell)\Gamma(2\ell)] \sim \ell$. Hence $\tphi$ is ill-defined and it is not a solution of the system. Note that, in order to discuss asymptotic solutions, the limit $t\to\infty$ must be taken \emph{after} resumming the terms of the series, that is, once this is shown to converge.


\subsection{Super-accelerating case ($H_0>1$, $q=1$), $\phi=t^p$}

The $\ell$-th term in the series $\tphi$ reads
\ba
\B^\ell\phi &=&(-1)^\ell\sum_{n=0}^{\ell} b_{\ell,n} t^{p-2n},\label{eqsa}\\
b_{\ell,n} &\equiv& (3H_0)^{\ell-n}\left[\prod_{k=0}^{2n-1}(p-k)\right]\nonumber\\
&&\quad\times\sum_{d_0=0}^{\ell-n}\dots\sum_{d_n=0}^{\ell-n}\delta\left(\sum_{m=0}^n d_m-\ell+n\right)\nonumber\\
&&\quad\times\prod_{j=0}^{n}(p-2j)^{d_j},
\ea
where the prefactor in front of $t$ in the last line is a polynomial in $(p-2j)$, $j=0\dots n$, of degree $\ell-n$ with all coefficients equal to 1. In the first line, when $n=0$ the product is equal to 1. The coefficients $b_{\ell,n}$ increase with $\ell$ and, again, $\tphi$ is not well-defined for $p\notin\mathbb{N}$. If $p\in\mathbb{N}$, the series \Eq{eqsa} terminates at $n_*=[p/2]$ (square brackets indicate the principal value of $p$: $p=2n_*+1$ or $p=2n_*$) and $\tphi$ is a finite sum of powers of $t$, odd (even) if $p$ is odd (even). In the case $p=1$, for instance [4D solution for $n=2$ in Eq.~\Eq{aexp}], 
\be
\tphi= \rme^{-3H_0r_*} t\propto \phi\,.
\ee


\subsection{Accelerating case ($H_0>1$, $q=-1$), $\phi=t^p$ and $\phi=\ln t$}\label{acc}

The first slow-roll parameter is constant, $\epsilon=H_0^{-1}$, and
\ba
\B^\ell\phi &=&(-1)^\ell\left[\prod_{n=0}^{\ell-1}(p-2n)(p-2n-1+3H_0)\right] t^{p-2\ell}\nonumber\\
&\equiv& b_{\ell}\, t^{p-2\ell}.\label{ser}
\ea
If $p\in\mathbb{R}\setminus\mathbb{N}^+$, the coefficients of the series defining $\tphi$ increase with $\ell$:
\ba
\lim_{\ell\to\infty}\left|\frac{c_{\ell+1}\B^{\ell+1}\phi}{c_\ell\B^{\ell}\phi}\right|&=&\lim_{\ell\to\infty} |c_1(p-2\ell) \nonumber\\
&&\quad\times(p-2\ell-1+3H_0)|\frac{t^{-2}}{\ell+1}\nonumber\\
&=&+\infty\,,\nonumber
\ea
for finite $H_0$ and $t$. A possibility to avoid the convergence problem is that $p$ is a positive even number, so that the series ends at $\ell_*\equiv p/2$. Then 
\be
\tphi=\sum_{\ell=0}^{p/2}c_\ell b_\ell t^{p-2\ell}
\ee
is constant at early times, $\tphi\sim t^p$ at late times, and the system is local (all the other functions $\cO_i$ and $U'$ are finite sums of powers of time). Another case where the series is finite is when
\be\label{cnd}
p-1+3H_0=2n\,,
\ee
for some natural $n$. 

The discussion easily extends to a more general Hubble parameter, $q\neq \pm 1$ in Eq.~\Eq{hq}. We can immediately discard also the case $\phi=\ln t$ since $\tphi$ would not converge:
\ba
\B^\ell\phi &=&(3H_0-1)(-1)^\ell\left[\prod_{n=0}^{\ell-2}2(1+n)(2n+3-3H_0)\right]\nonumber\\
&&\quad\times t^{-2\ell}\,,\qquad \ell>0\,.\label{lnser}
\ea
The series is finite only when $H_0=(2n+3)/3$, $n\in\mathbb{N}$.


\subsection{Preliminary assessment}

One may be tempted to conclude that either \textbf{(a)} the exact cosmological solutions described by Eq.~\Eq{hq} cannot be used as a base for nonlocal functions, or \textbf{(b)} this \emph{Ansatz} is viable only as long as the series defining $\tphi$ can be truncated at a finite order, in which case the system is trivially local at late times. 

As general statements, both (a) and (b) are proven to be wrong (also on Minkowski \cite{roll}). Although the definition Eq.~\Eq{lb} of the exponential kinetic operator is unique and unambiguous, what is not defined is the nonlocal solution \emph{expressed as an infinite series of powers of the d'Alembertian}. We shall see in Sec.~\ref{compa} that $\tphi$ defined in terms of the Green functions of the $\B$ operator not only reproduces the special cases of finite series in $t$, but is well-defined even when the above approach fails.

The reason is the following. We shall write an integral representation for $\tphi$ which allows us to sum the exponential series consistently. This solution satisfies by construction a diffusion equation which renders all equations of motion local in time. Instead of having an infinite number of differential equations, one reduces the problem to a set of algebraic equations in the parameters of the system.


\section{Nonlocal solutions: Integral representation}\label{strat}

We propose a new recipe to construct nonlocal solutions. We interpret $r_*$ as a fixed value of an auxiliary evolution variable $r$, so that the scalar field $\phi(r,t)$ is thought to live in $1+1$ dimensions (in this discussion the 3 spatial dimensions play no r\^ole). The first step is to find a solution of the corresponding local system, that is with $r_*=0$. This can be done by solving Eq.~\Eq{teom} or \Eq{teom2} with $r_*=0$, denoting the local homogeneous solution $\phi_{\rm loc}(t)=\phi(r_*=0,t)$ and considering this as the initial condition for a system that evolves in $r$ through the diffusion equation
\be\label{dif}
\a\p_r\psi(r,t)=\B\psi(r,t)\,, \qquad \psi(0,t)=\phi_{\rm loc}(t)\,,
\ee
for some (unknown) parameter $\a$. The solution of Eq.~\Eq{dif} is
\be\label{soldif}
\psi(r,t)=\rme^{(r/\a)\B}\phi_{\rm loc}(t)\,.
\ee
The action of nonlocal operators of the type $\rme^{qr\B}$ now becomes local for systems that solve Eq.~\Eq{dif}. In fact,
\be\label{tra}
\rme^{qr\B}\psi(r,t)= \rme^{\a qr\,\p_r} \psi(r,t)=\psi((1+\a q)r,t)\,,
\ee
and all the effect of the operator is in the shift of the auxiliary variable $r$. The function $\psi(s,t)$, $s\neq r$, satisfies a new diffusion equation, given by Eq.~\Eq{dif} with $r\to s$ everywhere (including the differential operator).

Since $r$ and $\p_r$ do not commute, we need an ordering prescription for the exponential in Eq. \Eq{tra}. We adopt the most natural one, setting all the derivatives $\p_r$ to the right of the powers of $r$, in such a way that the translation property Eq.~\Eq{tra} holds also in this case. In addition, this is the only prescription compatible with the diffusion equation \Eq{dif}. In fact,
\ba
\rme^{r\B}\psi(r,t)&=&\sum_{k=0}^\infty \frac{r^k }{k!}\B^k \psi(r,t)
=\sum_{k=0}^\infty \frac{(\a r)^k }{k!}\p_r^k \psi(r,t)\nonumber\\
&=&\psi((1+\a)r, t)\,.\label{tra2}
\ea
In view of the above discussion, we can look for solutions of the nonlocal equations of the type $\phi(r,t)=C_\psi(r) \psi(r,t)$, where $C_\psi(r)$ is a constant in time. Since $\phi_{\rm loc}$ is a solution of the local system ($r_*=0$), then
\be\label{cpsi}
C_\psi(r=0)=1,
\ee
and we can choose $C_\psi=\rme^{\b r}$, for some unknown $\b$. This is equivalent to allowing for a linear term in the diffusion equation for $\phi(r,t)$:
\be\label{difphi}
\a\,\p_r\phi(r,t)=\a\b\,\phi(r,t)+\B\phi(r,t)\,,
\ee
and the shift property \Eq{tra} is unaltered. Equation \Eq{teom2} becomes
\be\label{npeom}
(\B-m^2) \psi((\gamma-\a)r,t) = \rme^{-\b r}  \tilde U'[\rme^{\b r} \psi((\gamma+\a)r,t)]\,,
\ee
where we started from a configuration $\phi(\gamma r,t)$, absorbed a factor $\gamma$ in $\beta$, and used
\be
\tphi(\gamma r,t)=\rme^{\b r} \psi((\gamma+\a)r,t)=\rme^{-\a\b r} \phi((\gamma+\a)r,t)\,.
\ee
The parameter $\gamma$ is not constrained dynamically like $\a$ and $\b$; it is introduced only to avoid any ambiguity in the present notation. One can fix it, for instance, so that $\gamma=1+\a$ and $\phi(r\neq 0,t)$ on the left-hand side of Eq.~\Eq{npeom}.

In general, the normalization constant $C_U$ in $U$ will depend on $r$, so there is also the freedom to choose a function $C_U(r)$ such that $C_U(0)$ matches with the normalization of the local potential.

It should be stressed again that Eq.~\Eq{npeom} is local in $t$, at variance with other methods developed in the literature and based on a finite truncation of the exponential defining the nonlocal operator \cite{are04,AJ,AK}. This property is very important, as it permits to study the nonlocal equations of motion in limited time intervals. Moreover, to understand the asymptotic (in time) dynamical behaviour it is mandatory to look for solutions that, although approximate, sum the \emph{whole} exponential series, rather than to find solutions at any fixed truncation level $\B^\ell$ of the exponential operator. For instance, in Minkowski spacetime a fixed truncation of the spike solution \cite{roll} gives rise to spurious wild oscillations near the origin, which are absent in the full representation.

Finally, using Eqs.~\Eq{difphi} and \Eq{npeom}, the functions $\cO_1$ and $\cO_2$ become (here $\gamma=1+\a$)
\ba
\cO_1 &=&\a r\, \rme^{2r\b}\int_0^1 \rmd s\,\p_{\zeta r}\psi(\zeta r,\,t)\,(\a\p_{\xi r}-m^2)\psi(\xi r,\,t)\,,\nonumber\\\label{O1tr}\\
\cO_2 &=& \frac{2r}{\dot{\psi}^2((1+\a)r,\,t)}\int_0^1 \rmd s\,\dot\psi(\zeta r,\,t)\,(\a\p_{\xi r}-m^2)\nonumber\\
&&\qquad \times\,\,\dot\psi(\xi r,\,t)\,,\label{O2tr}
\ea
where $\zeta\equiv 1+\a(2-s)$ and $\xi\equiv 1+\a s$.

The calculation of $\phi(r,t)$ proceeds as follows.
\begin{enumerate}
\item[(i)] \emph{Find the eigenstates of the d'Alembertian operator}. First, we solve the equation
\be\label{eiga}
\B G_i(\mu,t) =\mu^2 G_i(\mu,t)\,,
\ee
that is,
\be\label{eig2}
\ddot G_i+3H\dot G_i+\mu^2 G_i=0\,,
\ee
where $\mu\in \mathbb{C}$ is either real or pure imaginary in order for $\B$ to be a self-adjoint operator. The eigenvalue equation is second-order and hence $i=1,\,2$.
\item[(ii)] \emph{Write the nonlocal solution as an expansion in the basis of eigenstates of $\B$.}
The problem then is to find a function $f_i(\mu)$ for each Green function $G_i$ such that the local solution $\phi(0,t)$ can be expressed as an integral transform,
\ba
\phi(0,t) &=&\int \rmd\mu\, [C_1 G_1(\mu,t)\,f_1(\mu)\nonumber\\
&&\qquad+C_2\,G_2(\mu,t)\,f_2(\mu)]\,,\label{phs}
\ea
while the localized field is a Gabor transform\footnote{Strictly speaking, a Gabor transform is realized when the $G_i$ are combinations of phases, which is the case of the Minkowski CSFT tachyon.}
\ba
\psi(r,t) &=&\int \rmd\mu\, \rme^{r\mu^2/\a}[C_1 G_1(\mu,t)\,f_1(\mu)\nonumber\\
&&\qquad+C_2\,G_2(\mu,t)\,f_2(\mu)]\,,\label{tphs}
\ea
for some constants $C_i$, $\a$ and $\b$. These integrals become series if $\mu$ varies on a discrete set. The advantage of the integral representation \Eq{tphs} is that the properties given by Eqs.~\Eq{dif} and \Eq{tra} are manifest. In other words, any function $\psi$ which can be represented as a convergent integral of the form \Eq{tphs} obeys the diffusion equation.
\item[(iii)] Verify that there exist some $C_i$, $C_U$, $\a$, and $\b$ such that Eq.~\Eq{tphs} is a solution of the equations of motion for $r=r_*$. This can be done in several ways, either by variational methods, or by expanding the equations in power series of $r_*$ or $t$ and imposing the coefficients of the expansions to solve the equations. 
\end{enumerate}
The two approaches discussed in Secs.~\ref{covan} and \ref{series} (covariant \emph{Ansatz} and direct construction) are special subcases of this analysis.

The heat kernel method of \cite{roll} would seem more direct than the one based on the Green functions of the d'Alembertian operator. However, on curved backgrounds, the problem of finding a kernel satisfying a heat equation with covariant d'Alembertian is much less trivial than in the Minkowski case. While on Minkowski spacetime such a kernel can be immediately guessed, on a FRW spacetime with a specific Hubble parameter one cannot proceed blindfolded and a systematic treatment is mandatory. An expansion of the tachyon field in terms of the $\B$ eigenfunctions still yields the desired result without extra computational effort.

We shall call \emph{local} the system of equations solved by $\phi_{\rm loc}(t)$ ($r=r_*=0$ everywhere), while the $(1+1)$-system solved by some $\phi(r,t)$ will be referred to as \emph{localized}. In phenomenological models, there is freedom in the choice of the nonlocal potential in a way which does not affect the local equations of motion. Equation \Eq{teom} can be written as
\be\label{teomh}
\B\phi=h(\phi)-h(\tphi)+U'(\tphi)\,,
\ee
where $h$ is an arbitrary function, while Eq.~\Eq{teom2} becomes
\be\label{teom2h}
\B \rme^{-2r_*\B}\tphi=\rme^{-r_*\B}[h(\rme^{-r_*\B}\tphi)-h(\tphi)]+\tilde U'.
\ee
When $h=0$ and the nonlocal potential is simply the local one with the substitutions $\phi\to \tphi$, $U_0\to C_U(r)$, we will call the system described by Eq.~\Eq{teomh} the minimal nonlocal system associated with the local one. The string case is nonminimal, $h(x)=m^2 x$. With this exception, there is no criterion for selecting $h$, and for the moment we will set it to vanish for simplicity ($m^2=0$). A nontrivial choice for $h$ may become relevant when looking for exact solutions. Another kind of extra term in the equations of motion can be of the form $r^n h(\phi,\tphi)$, $n>0$, which vanishes when $r\to 0$.

The equations of motion of the localized system are local in time. This fact has two crucial advantages. First, there is only one momentum conjugate to the scalar field and quantization is well-defined. Second, the equations of motion depend on a finite number of slow-roll parameters, and the slow-roll approximation is legitimate. The original nonlocal system contained infinitely many slow-roll parameters whose contribution could not be tested to be convergent; consequently, the slow-roll limit required much stronger assumptions out of analytic control \cite{cutac}.

On a de Sitter background, the two independent eigenfunctions are exponentials, 
$G_{1,2}=\rme^{\pm b_n t}$, where $b_n$ is given by Eq.~\Eq{bn} with $m$ replaced by $\mu$. Since exact and nontrivial cosmological solutions require a varying Hubble parameter, we move directly to these scenarios.


\section{Nonlocal braneworld system with $H=H_0 t^{-1}$}\label{1tbw}

In the power-law case, $H=H_0t^{-1}$, the independent solutions of the eigenvalue equation \Eq{eig2} are (\cite{GR}, formula 8.491.6)
\ba
&&G_1(\mu,t)=t^\nu J_{\nu}(\mu t)\,,\quad G_2(\mu,t)=t^\nu N_{\nu}(\mu t)\,,\\
&&\nu \equiv \frac12(1-3H_0)\,,
\ea
where $J_\nu$ and $N_\nu$ are the Bessel functions of the first and second kind.

Taking $C_2=0$, for $\mu$ real and positive we can write the scalar field as
\be\label{locmb}
\phi(0,t)= \int_0^{+\infty} \rmd\mu\, f_1(\mu)\, G_1(\mu,t)\equiv t^\nu F_\nu(t)\,,
\ee
where
\be
F_\nu(t)= \int_0^{+\infty} \rmd\mu\, f_1(\mu)\, J_\nu(\mu t)
\ee
is the \emph{Hankel transform} \cite{wikh}, whose inverse is
\ba
f_1(\mu) &=& \mu\int_0^{+\infty} \rmd t\, t\, F_\nu(t)\, J_\nu(\mu t)\nonumber\\
&=& \mu\int_0^{+\infty} \rmd t\, t^{1-\nu} \phi(0,t)\, J_\nu(\mu t)\,,\label{fmu}
\ea
for $\nu\geq -1/2$.

In the braneworld (BW) case, using formula 6.561.14 in \cite{GR},
\be\label{bwf}
f^{\rm (BW)}(\mu)=f_1(\mu) = \phi_0\, \frac{\Gamma(1+\theta/4)}{\Gamma(\nu-\theta/4)}\left(\frac{\mu}{2}\right)^{\nu-1-\theta/2}\,,
\ee
valid for $-4<\theta<-3H_0$ and $\nu-\theta/4\neq -n$, $n\in\mathbb{N}$. 

The integral \Eq{tphs} can be done exactly when $f_1(\mu)$ is Eq.~\Eq{bwf}. Using formula 6.631.1 of \cite{GR}, the function $\psi_1(r,t)$ constructed from $G_1$ is
\ba
\psi_1(r,t)&=& \phi_0\, 2^{1-\nu+\theta/2}\frac{\Gamma(1+\theta/4)}{\Gamma(\nu-\theta/4)}\, t^\nu\nonumber\\
&&\times \int_0^{+\infty} d\mu\,\mu^{\nu-1-\theta/2}\rme^{r\mu^2/\a}J_{\nu}(\mu t)\nonumber\\
&=& \phi_0\, \frac{\Gamma(1+\theta/4)}{\Gamma(\nu+1)}\,
\left(-\frac{4r}\a\right)^{\theta/4} \left(-\frac{\a t^2}{4r}\right)^{\nu}\nonumber\\
&&\quad\times\Phi\left(\nu-\frac{\theta}4;\,\nu+1;\,\frac{\a t^2}{4r}\right)\,,\label{got}
\ea
where $\Phi$ is Kummer's confluent hypergeometric function of the first kind, whose properties are summarized in the Appendix. The reader unfamiliar with it can just notice for the moment that it is defined as the series Eq.~\Eq{kumm}. The integration domain is the region $r/\a<0$ and $2\nu-\theta/2=1-3H_0-\theta/2>0$, but now one can analytically continue Eq.~\Eq{got} for every $H_0$, $r$, and $t$ (the typical values $\theta=\pm1$ do not constitute a problem). In order for the solution with $r/\a>0$ to be real, one must compensate the phase $\rme^{-\rmi\pi(\nu-\theta/4)}$ with the coefficient $C_1$ in front of it.

It is convenient to define
\be\label{zvar}
z \equiv \frac{\a t^2}{4r}\,,
\ee
so that
\ba
\psi_1(r,z) &=& \phi_0\, \frac{\Gamma(1+\theta/4)}{\Gamma(\nu+1)}\,\left(-\frac{4r}\a\right)^{\theta/4} (-z)^{\nu}\nonumber\\
&&\quad\times\Phi\left(\nu-\frac{\theta}4;\,\nu+1;\,z\right)\,.\label{good}
\ea
Equation \Eq{good} satisfies, by construction, the diffusion equation \Eq{dif} (see the Appendix). Notice that, for any $\theta$ and $z$, given $0<n,l\in\mathbb{N}$:
\begin{enumerate}
\item[(1)] $\nu-\theta/4\neq -n$ and $\nu+1\neq -l$: $\psi_1$ is a convergent series in $z$. If $\nu+1= -l$, the series Eq.~\Eq{kumm} starts at $k=l+1$.
\item[(2)] $\nu-\theta/4= -n$ and $\nu+1\neq -l$: $\psi_1$ is a polynomial of degree $\theta/4$ in $z$. This condition is given precisely by Eq.~\Eq{cnd} and it corresponds to one of the cases where the series representation of the $\rme^\B$ operator is finite and well-defined, in the sense that the local solution lives in the operator's domain. This is an example where the analytic continuation to the region $\nu-\theta/4<0$ is explicitly required.
\item[(3)] $-\theta/4=l-n+1$,\, $n>l$, $\nu$ noninteger: $\psi_1$ is an infinite hypergeometric series.
\item[(4)] $\nu-\theta/4= -n$,\, $\nu+1= -l$,\, $n>l$: The series \Eq{kumm} has only the terms $k=l+1,\dots,\,n$, with well-defined coefficients being $\Gamma(k-n)/\Gamma(-n)$ finite. Again $\psi_1$ is a polynomial of degree $\theta/4=n-l-1$ in $z$. This corresponds to $p=2,4,6,\dots$ in the notation of Sec.~\ref{series}.
\item[(5)] $\nu-\theta/4= -n$,\, $\nu+1= -l$,\, $n\leq l$: The Kummer function becomes an exponential times a polynomial of degree $l-n=-1-\theta/4$ in $z$, while the factor $\Gamma(n-l)/\Gamma(-l)$ in front of it is finite.
This case is unphysical ($\theta\leq-4$) and corresponds to $p=-2$ ($n=l$) and $p=-4,-6,-8,\dots$ $(n<l)$ in the notation of Sec.~\ref{series}.
\end{enumerate}
In the RS and GB braneworld cases, the last three instances are excluded.

Some asymptotic behaviours of Eq.~\Eq{good}:
\begin{itemize}
\item $z\to 0$ ($t\to 0$ or $r\to\infty$): 
\be\label{1t0bw}
\psi_1(r,z)\ \stackrel{z\to 0}{\sim}\  \phi_0\,\frac{\Gamma(1+\theta/4)}{\Gamma(\nu+1)}\, \left(-\frac{4r}\a\right)^{\theta/4}\,(-z)^\nu. 
\ee
The function diverges at the origin of time if $\nu<0$ ($H_0>1/3$).
\item $z\to \infty$ ($t\to \infty$ or $r\to 0$): Making use of the large $z$ expression \Eq{largez}, one has that, for $z\to +\infty$ ($r/\a>0$),
\ba
\psi_1(r,z)& \stackrel{z\to +\infty}{\sim}&  \phi_0\,\frac{\Gamma(1+\theta/4)}{\Gamma(\nu-\theta/4)}\,\left(-\frac{4r}\a\right)^{\theta/4}\nonumber\\
&&\qquad\times(-1)^\nu\,\rme^{z}\,z^{\nu+1+\theta/4}\,,\label{gau1}
\ea
while, for $z\to -\infty$ ($r/\a<0$), 
\ba
\psi_1(r,z) &\stackrel{z\to -\infty}{\sim}& \phi_0\left(-\frac{4r}{\a}\right)^{\theta/4} (-z)^{\theta/4}\nonumber\\ &=&\phi_0\,t^{\theta/2}\,.\label{gau2}
\ea 
\end{itemize}

A second, independent function which obeys the diffusion equation can be found directly from Eq.~\Eq{good}, instead of defining and calculating an invertible transform with the eigenstate $t^\nu N_\nu$.\footnote{We have done it as a consistency check. Since $G_2(\mu,t)=t^\nu N_{\nu}(\mu t)$, one can write $\phi(0,t)$ as the $N$-transform (\cite{bate}, formula 9.1)
\be\label{Ntran}
\phi(0,t)=\phi_0 t^{\theta/2}= \int_0^{+\infty} \rmd\mu\, f_2(\mu)\, G_2(\mu,t)\,,
\ee
where (\cite{bate}, formula 11.1)
\be
f_2(\mu) = \mu\int_0^{+\infty} \rmd t\, \phi(0,t)\, t^{1-\nu} {\bf H}_\nu (\mu t) = f_1(\mu) \tan(\pi\theta/4)\,,\nonumber
\ee
and ${\bf H}_\nu$ is the Struve function. The Gabor transform with the Neumann function $N_\nu$ turns out to be
\ba
\psi(r,t) &=& \int_0^{+\infty} \rmd\mu\, \rme^{r\mu^2/\a}\,f_2(\mu)\, G_2(\mu,t)\nonumber\\
          &=& \frac{\tan (\pi\theta/4)}{\tan(\pi\nu)}\,\psi_1(r,t)+\left\{1+\frac{\tan (\pi\theta/4)}{\tan[\pi(\nu-\theta/4)]}\right\}\,\psi_3(r,t)\,,\nonumber
\ea
where $\psi_3$ is defined below.} From what discussed in the Appendix, one can write $\psi_2$ as
\be
\psi_2(r,z) = \phi_0\,\left(\frac{4r}\a\right)^{\theta/4} \,\Psi\left(-\frac{\theta}4;\,1-\nu;\,z\right)\,,\label{good2}
\ee
where $\Psi$ is Kummer's confluent hypergeometric function of the second kind.
An eventual phase is reabsorbed into $C_2$. For any $r$ and $z$, given $0<n,l\in\mathbb{N}$:
\begin{enumerate}
\item[(1)] $\nu-\theta/4\neq -n$ and $\nu+1\neq -l$: $\psi_2$ is a convergent series in $z$.
\item[(2)] $\nu-\theta/4= -n$ and $\nu+1\neq -l$: $\psi_2$ is a polynomial of degree $\theta/4$ in $z$.
\item[(3)] $-\theta/4=l-n+1$,\, $n>l$, $\nu$ noninteger: $\psi_2$ is a polynomial of degree $\theta/4$ in $z$.
\item[(4)] $-\theta/4=l-n+1$,\, $\nu+1= -l$,\, $n>l$: $\psi_2$ collapses into $\psi_1$.
\item[(5)] $-\theta/4=l-n+1$,\, $\nu+1= -l$,\, $n\leq l$: $\psi_2$ is a polinomyal in $z$.
\end{enumerate}
Asymptotically:
\ba
&&\psi_2\ \stackrel{z\to 0}{\sim}\  \phi_0\,\left(\frac{4r}\a\right)^{\theta/4}\,\left[\frac{\Gamma(\nu)}{\Gamma(\nu-\theta/4)}+\frac{\Gamma(-\nu)}{\Gamma(-\theta/4)}\,z^\nu\right]\,,\nonumber\\\label{2z0}\\
&&\psi_2\ \stackrel{z\to-\infty}{\sim}\ \phi_0\, \rme^{\rmi \pi\theta/2}\, t^{\theta/2}\,,\label{2infy-}\\
&&\psi_2\ \stackrel{z\to+\infty}{\sim}\ \phi_0\, t^{\theta/2}\,.\label{2infy+}
\ea
According to the value of $\nu$, one should add next-to-leading order terms in Eq.~\Eq{2z0} to get a consistent expansion in $z$. At large $z$ (small $r$), one recovers the local solution for a suitable coefficient $C_2$ depending on the sign of $z$ (for general values of its first two arguments, $\Psi$ has a branch point at $r=0$). Modulo a phase, the local solution $\phi\propto t^{\theta/2}$ is manifestly the initial condition at $r=0$ of $\psi_1$ and $\psi_2$.


\subsection{Comparison with the series calculation}\label{compa}

Before looking for a solution of the nonlocal equations of motion, we compare the function
\be
\psi\equiv C_1\psi_1+C_2\psi_2\,,
\ee
found via the eigenfunction expansion method, with the brute-force series computed in Sec.~\ref{acc}. First, we check that it is always possible to choose a suitable linear combination of Eqs.~\Eq{good} and \Eq{good2} reproducing the series when this is finite (that is, when $\psi$ is a polynomial). Then, for cases 2 to 5 ($\a=1$ in what follows):
\begin{enumerate}
\item[(2)] Both $\psi_1$ and $\psi_2$ are proportional to the polynomial found via the series computation.
\item[(3)] Fixing $C_1=0$, $\psi$ is proportional to the polynomial found via the series computation.
\item[(4)] Same as case 2.
\item[(5)] Same as case 3.
\end{enumerate}
Note that these properties hold for any $r$, including the analytic continuation into the $r>0$ region. Therefore, it is consistent to impose
\be
C_1=0\,,
\ee
and take only one solution.

\begin{figure}\begin{center}
{\psfrag{t}{$t$}
\includegraphics[width=8.6cm]{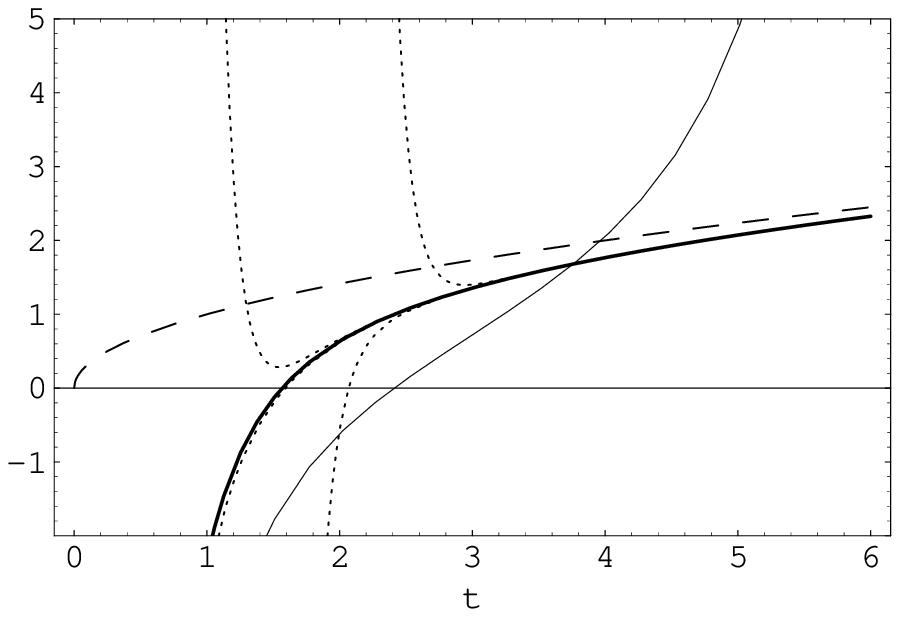}}
{\psfrag{t}{$t$}
\includegraphics[width=8.6cm]{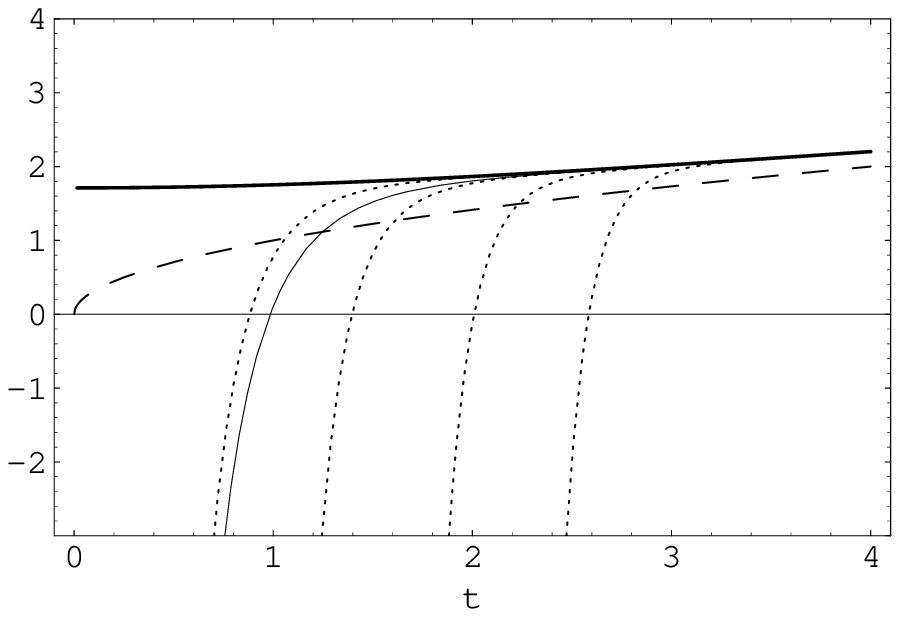}}
\caption{\label{fig1} The function $\psi(r,t)$ calculated via the series expansion and the eigenstates method. Here, $\theta=1$ (RS braneworld) and $\nu=-3/2$. The meaning of each dashed and dotted curve is explained in the text. In the upper panel ($r/\a=1$), the solid thin curve is $\psi_1$ with $C_1=\rme^{\rmi\pi(\nu-\theta/4)}\,$, while the solid thick curve is $\psi_2$ with $C_2=1$. In the lower panel ($r/\a=-1$), the solid thin curve is $\psi_1$ with $C_1=1$, while the solid thick curve is $\psi_3$ with $C_3=1$.}\end{center}
\end{figure}

\begin{figure}\begin{center}
{\psfrag{t}{$t$}
\includegraphics[width=8.6cm]{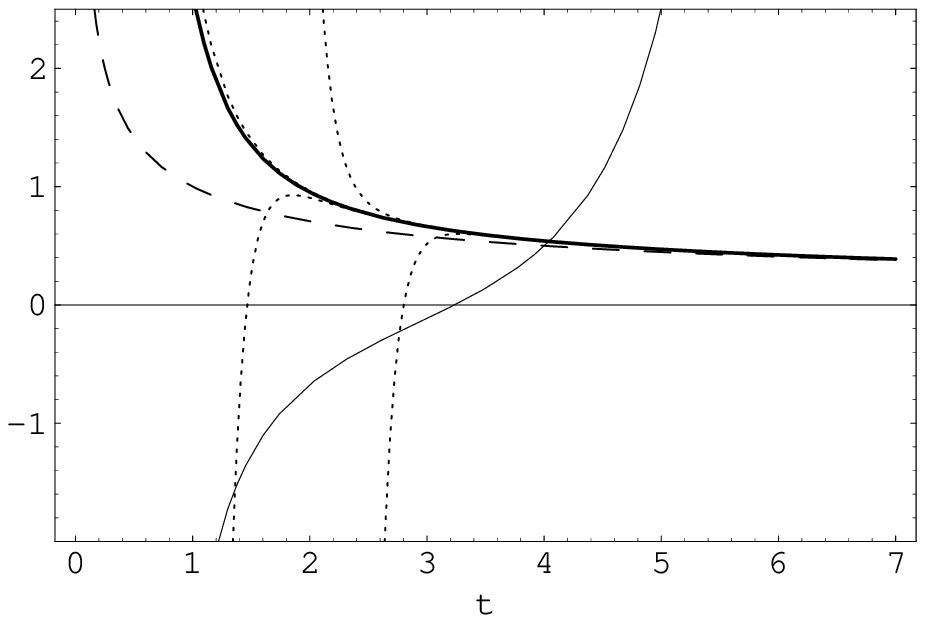}}
{\psfrag{t}{$t$}
\includegraphics[width=8.6cm]{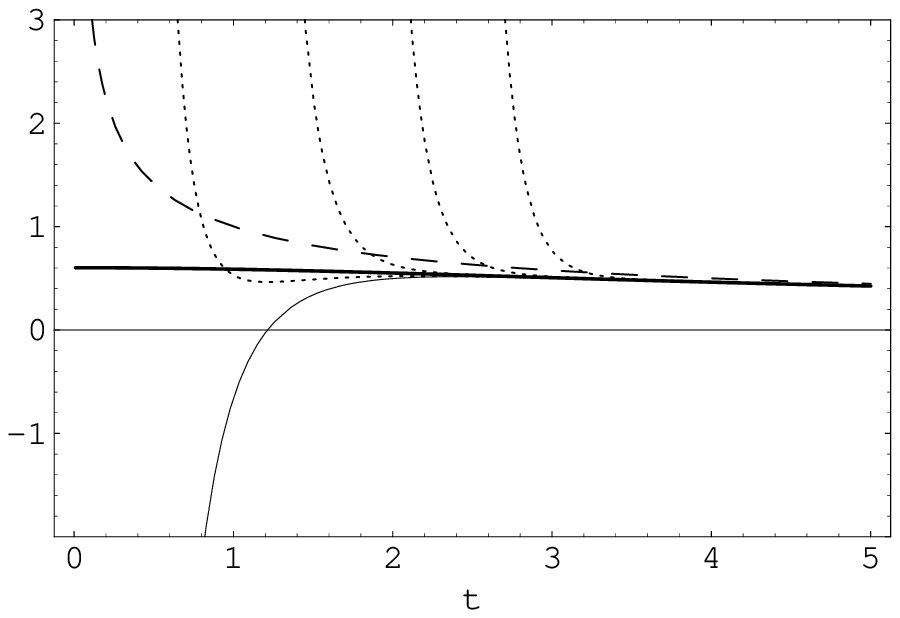}}
\caption{\label{fig2} The function $\psi(r,t)$ calculated via the series expansion and the eigenstates method. Here, $\theta=-1$ (GB braneworld) and $\nu=-3/2$. The meaning of each dashed and dotted curve is explained in the text. 
In the upper panel ($r/\a=1$), the solid thin curve is $\psi_1$ with $C_1=\rme^{\rmi\pi(\nu-\theta/4)}\,$, while the solid thick curve is $\psi_2$ with $C_2=1$. In the lower panel ($r/\a=-1$), the solid thin curve is $\psi_1$ with $C_1=1$, while the solid thick curve is $\psi_3$ with $C_3=1$.}\end{center}
\end{figure}

In case 1 ($-\theta/4\neq l-n+1$ and $\nu+1\neq -l$), for generic values of $\theta$ and $\nu$, the series built with terms as in Eq.~\Eq{ser} is not convergent, while both $\psi_1$ and $\psi_2$ are well-defined functions of time. When $z<0$, $\psi_2$ is complex-valued for general $\nu$. To get a real solution, there are two options: to set $C_2=0$ or
\be\label{caste}
C_1=C_2\,\rme^{\rmi\pi(\nu-\theta/4)}\,\frac{\sin(-\pi\theta/4)}{\sin\pi\nu}\,.
\ee
It is worth recording this linear combination as some coefficient $C_3$ times
\ba
\psi_3(r,z)
&=&\phi_0\, \frac{\Gamma(1-\nu+\theta/4)}{\Gamma(1-\nu)}\, \left(-\frac{4r}\a\right)^{\theta/4}\nonumber\\
&&\qquad\times\Phi\left(-\frac{\theta}4;\,1-\nu;\,z\right)\,,\label{good3}
\ea
with asymptotic behaviour
\ba
&&\psi_3\ \stackrel{z\to 0}{\sim}\  \phi_0\,\frac{\Gamma(1-\nu+\theta/4)}{\Gamma(1-\nu)}\,\left(-\frac{4r}\a\right)^{\theta/4}\,,\label{3t0bw}\\
&&\psi_3\ \stackrel{z\to+\infty}{\sim}\ \phi_0\,\frac{\Gamma(1-\nu+\theta/4)}{\Gamma(1-\nu)}
\left(-\frac{4r}\a\right)^{\theta/4}\rme^{z}z^{\nu-1-\theta/4},\nonumber\\\\
&&\psi_3\ \stackrel{z\to-\infty}{\sim}\ \phi_0\,t^{\theta/2}\,.
\ea
Figs.~\ref{fig1} and \ref{fig2} show the function $\psi(r,t)$ as superposition of eigenstates of the d'Alembertian and as truncated series. The dashed line is the solution of the local system ($r=0$). The dotted curves are the series functions given by Eqs.~\Eq{lb} and \Eq{dres}, truncated at $\ell=2,5,8,11$ (from left to right). The solid curves are linear combinations of $\psi_1$ and $\psi_2$ as explained in the caption; these are the main result of this part, to be compared with the truncated series and the local case.
To summarize:
\begin{enumerate}
\item[(i)] Case 1 ($\nu-\theta/4,\nu\notin\mathbb{Z}^-$): For $r/\a>0$, only $\psi_2$ reproduces the series computation at large $t$, while for $r/\a<0$ only a real linear combination of $\psi_1$ and $\psi_3$ is allowed. Note that the boundary condition given by the local solution of the local system is automatically satisfied.
\item[(ii)] Cases 2 to 5 (finite series): Only $\psi_2$, for any $r/\a$, is required. In case 2, being the physical one, $\psi_1\propto\psi_2$. 
\item[(iii)] In \emph{all} cases, the heat equation is fulfilled.
\end{enumerate}


\subsection{Solutions of the equations of motion}\label{solutionBW}

The problem of finding solutions of the scalar nonlocal equation with the local solution as boundary condition gets simplified if one takes into account only one of the three functions $\psi_i$ discussed above. Considering $z$ and $r$ as independent variables and using the diffusion equation, the equation of motion for the scalar field is now first-order for solutions of the form $\psi(r,z)=r^{-b} u(z)$ for some $b$. Actually,
\be
\B \psi=\frac\a{r}(r\p_r-z\p_z)\psi=-\frac\a{r}(b+z\p_z)\psi\,.
\ee
Eqs.~\Eq{fophi}, \Eq{fopsi}, and \Eq{beo} yield
\be
\psi_1^+(r,\,t)=
-\frac{\theta r}{\a}\, C_U(r)\, \rme^{-4\b r/\theta}\left[\psi_1\left((1+2\a)r,\,t\right)\right]^{1-4/\theta},\label{exeo1}
\ee
and
\be
\psi_2^+(r,\,t) =
\frac{4r}{\a}\, C_U(r)\, \rme^{-4\b r/\theta}\left[\psi_2\left((1+2\a)r,\,t\right)\right]^{1-4/\theta}\,,\label{exeo2}
\ee
while for $\psi_3$
\ba
\psi_3^+(r,\,t) &=&
\frac{r}{\a}(4\nu-\theta)\, C_U(r)\, \rme^{-4\b r/\theta}\nonumber\\
&&\qquad\times\left[\psi_3\left((1+2\a)r,\,t\right)\right]^{1-4/\theta}\,,\label{exeo3}
\ea
where the function $C_U(r)$ is arbitrary except for the property $C_U\to 1$ as $r\to 0$, and the superscript $+$ indicates that the first argument of the Kummer function in $\psi_i$ is translated by $+1$. Case 1 with $r/\a>0$ and case 2 for any $r/\a$ are described by Eq.~\Eq{exeo2}, while for case 1 with $r/\a<0$ one can choose between Eqs.~\Eq{exeo1} and \Eq{exeo3}. Case 1 with $r/\a<0$ also admits a linear combination of $\psi_1$ and $\psi_2$.

Here one can clearly see the nature of the Cauchy problem in nonlocal systems. These equations are purely algebraic in the scalar field. To solve the original equations of motion, one should find an infinite set of initial conditions $\phi_0$, $\dot\phi_0$, $\ddot\phi_0$, \dots\, . If $\phi$ is analytic, this is equivalent to find the Taylor expansion of $\phi$ around a point, and to know the initial conditions means to already know the solution \cite{MZ}. In the above expressions, the set of initial conditions has been reduced to a \emph{finite} one given by the parameters $\phi_0,\,H_0,\,\dots$ of the solution $\phi$, which is known in its functional form thanks to the heat equation.

In the original equation of motion for $\phi$ there appeared all the derivatives of the field, that is, the full tower of slow-roll parameter. While in local systems there are only a finite number of slow-roll parameters and all these can be set to be small, in the nonlocal case it is not guaranteed \emph{a priori} that the slow-roll tower as a whole is negligible with respect to $O(1)$ terms; the results of Sec.~\ref{compa} are a demonstration of this fact. In \cite{cutac} an extra condition for this approximation to be true was given, but here we do not even need it. In fact, $\phi(r,t)$ depends only on the first slow-roll parameter $\epsilon=3/(1-2\nu)$, and the standard slow-roll approximation can be safely applied. This is a nontrivial property which may justify \emph{a posteriori} the truncation scheme sometimes adopted in the past \cite{are04,AJ,AK} in the case when the \emph{full} nonlocal solution is known.

Let us now find solutions of the dynamical equations. First, we consider Eq.~\Eq{exeo2}. An exact solution can be inferred by looking at the asymptotic behaviour of $\psi_2$; expansions in $z$ (that is, $t$ or $r$) are always well-defined and there is no issue of convergence (although there may not be a global solution of the form $\phi(r,t)$).
At large $z$, matching powers of $z$ cancel out and at lowest order [$M=0$ in Eq.~\Eq{psidive}] one gets
\be
C_U(r)=\rme^{4\b r/\theta}\,.
\ee
The scale $\beta$ is always fixed if $C_U(r_*)$ is given from the beginning or, as alternative, $C_U$ defines the shape of the renormalization factor in front of $\psi(r,t)$. At next-to-leading order ($M=1$), the equation of motion is safistied for $\a=2/(4\nu-\theta)$, while at third order ($M=2$) one finds the conditions $\nu=1$ (describing a contracting universe, which we will ignore) or
\be\label{condi1}
\frac{\theta}4-\nu=1\quad\Rightarrow\quad \a=-\frac12\,,
\ee
giving a \textbf{case 2} exact solution of Eq.~\Eq{exeo2} in high-energy braneworld power-law cosmologies:
\be\label{solution2}
\psi_2(r,\,t) = \phi_0 t^{\theta/2}\left(1+\frac{2r\theta}{t^2}\right)\,.
\ee
In the Randall--Sundrum braneworld ($\theta=1$), $\nu=-3/4\Rightarrow H_0=1-\theta/6=5/6<1$, which is outside the accelerating regime. In the Gauss--Bonnet case, $\nu=-5/4\Rightarrow H_0=1-\theta/6=7/6>1$ which gives very mild acceleration. This solution is steeper than the local one near the origin, as one can also see by considering the expansion around $z=0$ of $\psi_2$, taking $-\nu\gg 1$ and noticing that powers of $z$ do not match in the left- and right-hand side of Eq.~\Eq{exeo2}. The sign of $r\theta$ determines whether $\phi\to \pm\infty$ as $t\to 0$; if one wants to stick to the $r>0$ case, the GB $\phi$ is an increasing function which vanishes at $t_*=\sqrt{2 r}$.

The value $\a=-1/2$ may seem problematic because the local function with $r=0$ in the right-hand side of Eq.~\Eq{exeo2} would be equated to a nonlocal part with $r\neq 0$. This is only a matter of notation. In our formalism, $r$ can be translated arbitrarily via $\rme^\B$ operators, and the equation of motion remains nonlocal regardless what translation is chosen to represent the $r$ dependence. The special value $\a=-1/2$ defines a ``coordinate system'' where $\tphi\propto \phi_{\rm loc}$. 

The field $\phi$ in the Friedmann equation is $\rme^{\b r}\psi_2(r/2,\,t)$. When $\a=-1/2$, Eqs. \Eq{O1tr} and \Eq{O2tr} become ($m^2=0$)
\ba
\cO_1 &=& -r\, \rme^{2\b r}\int_0^1 \rmd s\,\p_{sr}\psi(sr/2,\,t)\,\p_{sr}\psi((1-s/2)r,\,t)\,,\nonumber\\\\
\cO_2 &=& \frac{2r}{\dot{\psi}^2(r/2,\,t)}\int_0^1 \rmd s\,\dot\psi(sr/2,\,t)\,\p_{sr}\dot\psi((1-s/2)r,\,t)\,.\nonumber\\
\ea
Equation \Eq{solution2} is linear in $r$ and the computation of these quantities is straightforward:
\be
\cO_1 \propto t^{\theta-4}\,,\qquad \cO_2 = \frac{t^4}{[t^2+(\theta-4)r]^2}-1\,.
\ee
The Friedmann equation is $f_\textsc{frw}(t)\sim t^{\theta-2}[O(1)+O(t^{-2})+O(t^{-4})]$ and is not solved exactly, since the $\cO_i$ do not match with the the rest of the equation, $\sim t^{\theta-2}$. It is possible to choose $\b$ such that the $O(1)$ term in the Friedmann equation vanishes, but the other terms cannot be cancelled. At large $t$ the $\cO_i$ are subdominant and the usual local cosmology is consistently recovered. It is confirmed \cite{are04,cutac} that, in principle, the sign of the kinetic term of $\phi$ can change during the evolution of the universe (at early times, $\cO_2\sim -1$, while at late times $\cO_2\sim 0$). For instance, if $r<0$, $\text{sgn}(1-\cO_2)>0$ always; if $r>0$, $\text{sgn}(1-\cO_2)<0$ (and actually blows up) for a finite period during the evolution. However, this qualitative result is not robust because, at times when $\text{sgn}(1-\cO_2)<0$, the Friedmann equation is not solved with a satisfactory level of accuracy, $|f_\textsc{frw}(t)/(H^{2-\theta}+\beta_\theta^2\rho)|\gtrsim 10\%$.

Equation \Eq{solution2} is also the only asymptotic solution of the scalar equation at early times.\footnote{The reader can check this by expanding the equation of motion with $\a=-1/2$ for $t\ll 1$. $\psi_1$ collapses into the same function, while $\psi_3\sim \text{const} +O(t^2)$. One can absorb the constant in the definition of $\psi_3$ but it is not possible to match the next-to-leading term in the right-hand side of the equation of motion.} Since it is not an early-time solution of the Friedmann equation, there are no nontrivial solutions of the nonlocal system at early times.

To get solutions falling in \textbf{case 1}, one has to consider more general values of $\nu$. For instance, $-\nu\gg 1$ corresponds to slowly rolling fields in an expanding universe. We have checked for all $\psi_i$ and several values of the parameters that none of equations of motion is exactly solved, and the only regime where they are solved approximately is at late times, that is, when $\phi(r,t)$ tends to the local solution $t^{\theta/2}$. To be more precise, the equation of motion for the scalar field is of the form $A=B$, where $A$ and $B$ are noncomposite functions of time. One can choose a time $t_*$ above which $|(A-B)/(A+B)|<10^{-n}$ for some given $n$. A way to get approximate (and still almost local) expanding solutions at early times is to consider $\nu>0$ and then perform a time reversal of the equations. If $\nu-\theta/4$ is chosen to be a positive integer, $\phi(r,t)$ can be written in terms of exponentials, powers and incomplete gamma functions.

We note that the higher the value of $|\nu|$, the higher the characteristic time $t_*$. This important feature is compatible with the notion that it is difficult to realize inflation in nonlocal theories, with the possible exception of the special case, described in Sec.~\ref{covan}, where $\B\phi\propto\phi$ \cite{BBC,lid07}. If analytic, the solution of the equations of motion is uniquely determined by the infinite set of initial conditions one should specify in the nonlocal Cauchy problem. As we have seen, the latter can be conveniently translated into a finite set of algebraic constraints, which fixes most or all of the parameters of the theory. The solution is very rigid and not all values of the parameters are allowed; it firmly holds in memory the initial conditions, and any mechanism (such as inflation) which tends to wipe them out is unlikely to be realized.

The condition $\a=-1/2$ simplifies the calculations because in that case the potential term is a monomial, but we have verified that all these results can be generalized, with no appreciable changes, to other values of $\a$.


\section{Nonlocal 4D system with $H=H_0 t^{-1}$}\label{1t4d}

In four dimensions ($\phi=\phi_0 \ln t$), using formula 6.771 in \cite{GR}, the kernel realizing the integral transform Eq.~\Eq{locmb} is
\ba
f^{\rm (4D)}(\mu)&=&f_1(\mu)\nonumber\\
&=& \frac{\phi_0}{2\Gamma(\nu)}\,\left(\frac{\mu}2\right)^{\nu-1} \left[\psi(\nu)+\psi(1)-\ln\left(\frac{\mu}2\right)^2\right]\,,\nonumber\\\label{fmuj}
\ea
where $\psi=\p_x \ln \Gamma(x)$ is the digamma function and $\gamma=-\psi(1)\approx -0.577$ is the Euler--Mascheroni constant. The integral \Eq{fmu} is done in the region of the parameter space $\mu>0$, $\nu>1/2$, implying $H_0<0$. One can either analytically continue Eq.~\Eq{fmuj} to the region $\nu\leq 1/2$ or, if interested in post-big bang cosmology, make a time reversal to get a superaccelerating solution.

The resulting nonlocal function can be calculated directly and reads
\ba
\psi_1 &=& \phi_0\, \frac{(-z)^\nu}{2\Gamma(\nu+1)}\nonumber\\
&&\times\left\{\left[\psi(1)+\frac12\ln\left(\frac{4r}\a\right)^2\right]\,\Phi\left(\nu;\,1+\nu;\,z\right)\right.
\nonumber\\
&&\qquad\left.+\nu\sum_{k=1}^{+\infty} \frac{\psi(\nu)-\psi(\nu+k)}{\nu+k}\,\frac{z^k}{k!}\right\}\label{4d1}\\
&=& \phi_0\, \frac{(-z)^\nu}{2\Gamma(\nu)} \sum_{k=0}^{+\infty}\frac{z^k}{k!}\nonumber\\
&&\times \frac{\psi(1)+\tfrac12\ln\left(\tfrac{4r}\a\right)^2+\psi(\nu)-\psi(\nu+k)}{\nu+k}\,,
\ea
which is convergent with infinite radius of convergence. When $\nu \in \mathbb{Z}^-$, the series terminates at $k=k_{\rm max}=-\nu$. 

Here we point out a trick which allows to get the same result with slightly less computational effort (in fact, the direct calculation relies on a similar shortcut). One should treat the 4D and braneworld cases separately because, setting $\theta=0$ in Eq.~\Eq{bwf}, $f^{\rm (BW)}\big|_{\theta=0}\neq f^{\rm (4D)}$. This is expected, since the local braneworld solution $\phi\propto t^{\theta/2}$ is trivial for $\theta=0$. However, the following relation between the local solutions is valid:
\be
\left.\frac{\p}{\p\theta}[\phi_0^{\rm (BW)}t^{\theta/2}]\right|_{\theta=0}=\frac{\phi_0^{\rm (BW)}}{2\phi_0^{\rm (4D)}}\,[\phi_0^{\rm (4D)}\ln t]\,,
\ee
where we have treated $\phi_0^{\rm (BW)}$ as a fixed constant, and one can verify that
\be\label{resca}
f^{\rm (4D)}(\mu)=\frac{2\phi_0^{\rm (4D)}}{\phi_0^{\rm (BW)}}\left.\frac{\p f^{\rm (BW)}(\mu)}{\p\theta}\right|_{\theta=0}\,.
\ee
Then one can compute $\phi(r,t)$ in the simpler case of the braneworld and then get the nonlocal field in four dimensions by just a derivation. We need formula \Eq{deal}, yielding 
\ba
&&\p_\theta \Phi(\nu-\theta/4;\,1+\nu;\,z)\big|_{\theta=0} \nonumber\\
&&\qquad= -\frac{\nu}{4}\sum_{k=1}^{+\infty} \frac{\psi(\nu+k)-\psi(\nu)}{\nu+k}\,\frac{z^k}{k!}\,.\nonumber
\ea
Taking the derivative with respect to $\theta$ of Eq.~\Eq{good}, after the rescaling of Eq.~\Eq{resca} one gets exactly Eq.~\Eq{4d1}.

The asymptotic behaviours of $\psi_1$ can be calculated directly or by taking the $\theta$-derivative of Eqs.~\Eq{1t0bw}--\Eq{gau2}. In particular,
\ba
&&\psi_1\ \stackrel{z\to 0}{\sim}\  \frac{\phi_0}{2\Gamma(\nu+1)} \left[\psi(1)+\frac12\ln\left(\frac{4r}\a\right)^2\right](-z)^\nu\,,\nonumber\\
&&\psi_1\ \stackrel{z\to -\infty}{\sim}\ \phi_0\ln t\,,
\ea
while in the limit $z\to +\infty$, $\psi_1\propto \rme^z z^{\nu+1} (\text{const}+\ln z)$.

Although one can define an antitransform via a function $f_2(\mu)\propto \mu^{\nu-1}$, the integral transform Eq.~\Eq{Ntran} for $G_2$ is singular, as the parameter range lies outside the definition domain of the transform.

Nevertheless, in the braneworld case this transform exists, and the resulting functions $\psi_{2,3}$ are well-behaved. Another 4D nonlocal function can be found by deriving one of the other independent braneworld functions with respect to $\theta$. For instance
\ba
&&\p_\theta \Phi(-\theta/4;\,1-\nu;\,z)\big|_{\theta=0} \nonumber\\
&&\qquad= -\frac{\Gamma(1-\nu)}4\sum_{k=1}^{+\infty} \frac{\Gamma(k)}{\Gamma(1-\nu+k)}\,\frac{z^k}{k!}\nonumber\\
&&\qquad= \frac{z}{4(\nu-1)}\,\,_2F_2(1,\,1;\,2,\,2-\nu;\,z)\,,\nonumber
\ea
and one gets
\ba
\psi_3 &=& \frac{\phi_0}2\left[\psi(1-\nu)+\frac12\ln\left(\frac{4r}\a\right)^2\right.\nonumber\\
&&\quad\left.-\frac{z}{1-\nu}\,\,_2F_2(1,\,1;\,2,\,2-\nu;\,z)\right]\,,\label{4d2}
\ea
where $_2F_2$ is a generalized hypergeometric function defined in Eq.~\Eq{2F2}. $\psi_3$ is well-defined only when $\nu<1$. Asymptotically,
\ba
&&\psi_3\ \stackrel{z\to 0}{\sim}\  \frac{\phi_0}{2} \left[\psi(1-\nu)+\frac12\ln\left(\frac{4r}\a\right)^2\right]\,,\\
&&\psi_3\ \stackrel{z\to -\infty}{\sim}\ \phi_0\ln t\,,
\ea
while in the limit $z\to +\infty$, $\psi_3\propto \rme^z z^{\nu-1} (\text{const}-\ln z)$. It is checked that Eqs.~\Eq{4d1} and \Eq{4d2} obey the diffusion equation \Eq{dif}. 

A particular linear combination which can be derived from the braneworld function $\psi_2$ is
\be
\psi_2 = \psi_3-\frac{\phi_0}2\,\left[\frac{\pi}{\sin\pi\nu}+\Gamma(-\nu)(-z)^\nu\Phi(\nu;\,1+\nu;\,z)\right]\,.\label{d4d2}
\ee
This function differs from $\psi_3$ by a term which vanishes at late times, and generalizes the constant field shift parametrizing the class of exact local solutions.


\subsection{Comparison with the series calculation}\label{compa4d}

\begin{figure}\begin{center}
{\psfrag{t}{$t$}
\includegraphics[width=8.6cm]{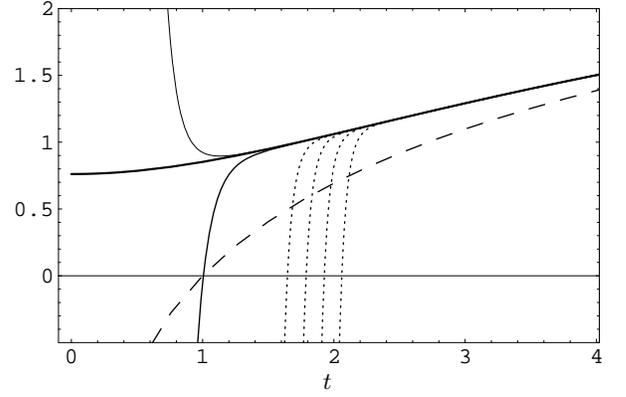}}
\caption{\label{fig6} The 4D function $\psi(r,t)$ calculated via the series expansion and the eigenstates method. The dashed curve is the local solution $\phi=\ln t$, while the dotted curves are the series-type functions at different level truncations. The solid curves are (increasing thickness) $\psi_1$, $\psi_2$, and $\psi_3$, respectively. Here, $\nu=-3.9$, $r=0.26$, $\a=-1$.}\end{center}
\end{figure}

Figure \ref{fig6} shows the functions $\psi_{i}(r,t)$ together with the divergent truncated series calculated before.
The dashed line is the solution of the local system ($r=0$). The dotted curves are the series function given by Eqs.~\Eq{lb}, \Eq{dres}, and \Eq{lnser}, truncated at $\ell=16,18,20,22$ (from left to right). The solid curves are $\psi_{1,2,3}$ as explained in the caption.

Again, it is possible to choose the parameter $\nu$ ($H_0$) in a region where the series function diverges while the functions $\psi_i(r,t)$ calculated via the diffusion equation are well-defined.


\subsection{Solutions of the equations of motion}\label{solution4d}

The local potential is $U(\phi)=U_0 \rme^{-2\phi/\phi_0}$, where $U_0=-\nu\phi_0^2$ and $\phi_0=\sqrt{2H_0/\kappa^2}$. Under the substitutions $\phi\to\tphi$ and $U_0\to U_0 C_U(r)$, the nonlocal equation of motion for the scalar is
\be\label{4deom1}
\B\psi_i(r,t)=  2\nu\,\rme^{-\b r}C_U(r) \exp[-2\rme^{\b r}\psi_i((1+2\a)r,t)]\,,
\ee
where we have cancelled the $\phi_0$ factors in the $\psi_i$'s. The function $C_U(r)$ is arbitrary except for the property $C_U\to 1$ as $r\to 0$. To get the local solution as an asymptotic solution at late times, it must be $C_U(r)=\rme^{\b r}$.

The left-hand side of the equation of motion can be worked out explicitly. It is uninstructively complicated for $\psi_1$, while for $\psi_2$ and $\psi_3$
\ba
\B\psi_2 &=& \frac{\a}{2r}\left(1-\rme^z\{z E_\nu(z)+\Gamma(1-\nu)[(-z)^\nu-z^{\nu}]\}\right)\,,\label{boxpsi2}\nonumber\\\\
\B\psi_3 &=& \frac{\a}{2r}\left\{1+\rme^z z^\nu [\Gamma(1-\nu)-\Gamma(1-\nu,\,z)]\right\}\,,
\ea
where $E_\nu(z)=\int_1^{\infty} \rmd s\,s^{-\nu} \rme^{-sz}$ is the exponential integral function and $\Gamma(1-\nu,\,z)=\int_z^\infty\rmd s\,s^{-\nu}\rme^{-s}$ is the incomplete gamma function. For $r/\a<0$, the first expression is real, while for $r/\a>0$ the definition \Eq{d4d2} needs an extra $(-1)^\nu$ factor in the last term, so that the term in square brackets in Eq.~\Eq{boxpsi2} vanishes.

In order to find asymptotic solutions for general $\nu$, one can choose $\a=-1/2$ so that the potential becomes a power of $t$ ($t^{-2\rme^{\b r}}$) which can be matched with the left-hand side of the equation. Assuming $\nu<0$, Eq.~\Eq{4deom1} for $\psi_1$ is satisfied at lowest order in $t\ll 1$ when
\ba
\rme^{\b r} &=& -\nu\,,\label{4dconst1}\\
C_U &=& \frac{1-\nu\psi(1)-\nu\ln (8r)}{(8r)^{1+\nu}\Gamma (1+\nu)}\,.
\ea
The sign of $C_U$ depends on the choice of $\nu$. For $\psi_2$, Eq.~\Eq{4dconst1} still holds together with
\be
C_U = -\frac{\Gamma (1-\nu)}{(8r)^{1+\nu}}\,.
\ee
In an accelerating regime with $r>0$, $C_U<0$. As regards $\psi_3$ at early times, at next-to-lowest order $\psi_3\sim -1/(4r) +O(t^2)$. One can absorb the constant in the definition of $\psi_3$ and eventually match the next-to-leading term. The latter step can be done after a generalization of the renormalization factor $\rme^{\b r}$ to a function $C_\psi(r)$ such that $C_\psi(r)\to1$ as $r\to 0$ and $C_\psi(r_*)<0$.

The Friedmann equation is not solved at early times by these choices of the parameters, and as in the braneworld case the only acceptable asymptotic solutions are those at late times.


\section{Discussion}\label{disc}


\subsection{Summary of the main results}

We have presented a method which allows us to explicitly work out nonlocal fields of nonlocal systems which do not necessarily admit a  representation as a series of $\B$ operators. This is the main difference relative to other proposals: it is not mandatory to truncate a series-type solution to calculate its coefficients, because the solution itself is already resummed. The simple case of power-law cosmology has been studied both in a pure four-dimensional scenario and on a brane with modified high-energy Friedmann equation. In all cases the found solutions are indeed approximate and close to the local behaviour.


\subsection{Comparison with other studies}

We want to compare our results with those recently appearing in literature. We already went through the subtleties of a truncation procedure \cite{are04,AJ,AK}, so we comment on the remaining points of view proposed on this stimulating subject.

In Ref.~\cite{AV,kos07}, the operator $\rme^\Box$ is represented as a Weierstrass product which allows one to recast the nonlocal tachyon Lagrangian as that of an infinite number of decoupled scalars $\psi_n$ which are eigenfunctions (with eigenvalue $\mu_n^2$) of the d'Alembertian. The latter is not a self-adjoint operator acting on these scalars, and the resulting total Hamiltonian has a continuous real spectrum which cannot describe particle modes.\footnote{See \cite{AV2} for a first step in the treatment of the quantization problem. Note that the $1+1$ Hamiltonian formalism avoids all difficulties associated with an infinite number of degrees of freedom; namely, there is only one local, physical field $\phi(r,t)$ to be quantized via a finite set of canonical commutation relations.} 

However, the $\B$ operator is assumed not to be self-adjoint because the authors look for kink-like solutions mediating from the unstable open-string vacuum to the stable closed-string vacuum at $t=+\infty$. Then, the characteristic equation for the differential operator cannot be solved for finite (real) eigenvalues $\mu_n^2$ in Eq.~\Eq{eiga}. As a consequence, solutions of this kind necessarily oscillate. Note that if one extends the approximate solutions of \cite{AK,kos07} also to negative times, these may have wild oscillations.

On phenomenological grounds, the author of \cite{kos07} considers a class of Lagrangians of the form ${\cal L}=\phi F(\B) \phi-\s\phi^{p+1}$, where $F$ is an unspecified complex operator. A notorious toy model falling into this class is the $p$-adic string \cite{FO,FW,BFOW}. The analysis assumes that $F(z)$ is analytic on a certain domain including the point $z=0$, so that $F$ admits Taylor expansion around the origin. These requirements translate into imposing that the operator $F(\B)$ can be written as a power series in $\B$ when acting on the physical states $\phi$. The field $\phi$ is written as a superposition of the scalars $\psi_n$, which have consequently the same r\^ole of the Green functions $G(\mu,t)$ of Eq.~\Eq{eiga}. Here we have considered only the case $\mu^2\in \mathbb{R}$, but it would be interesting to apply our formalism also to cases with $\Im (\mu^2)\neq 0$ and confront the results with the perturbative solutions of \cite{kos07}.

In \cite{AKV}, an action with a potential of the form \Eq{pot} with a quadratic nonlocal term is solved exactly on Minkowski space (via a Weierstrass product formulation) and its solutions of type $\sum_n C_n \rme^{\mu_n t}$ are classified.  On an FRW background, solutions are found under three different approximations: a truncated expansion (up to some $n_{\rm max}$) in terms of the local scalars $\psi_n$ which are also equipped with \emph{ad hoc} potentials; a truncation of the Weierstrass product; and a truncation of the Taylor series defining the kinetic nonlocal operator. Either method is effective only for computing finite (eventually large-order) series-type solutions; the order of the truncation determines the number of spurious solutions of the equations of motion (see also \cite{cutac} for a discussion on this point). Also, it is not clear whether the same treatment can lead to exact or well-defined approximated solutions in the case of more realistic potentials like those of effective SFT.

The $p$-adic string in a cosmological context is studied in \cite{BBC,BC}, where a relaxation of the $\eta$ problem is proposed (see also \cite{lid07}). Again, the considered solutions are finite series and it is conjectured that resummation of the kinetic operator is possible at arbitrary order. Here we gave negative examples with a different \emph{Ansatz} for $\phi$, so one cannot disprove this statement. Even if resummation were not possible, one could not conclude that this type of solutions is not viable; according to our perspective, it would not belong to the spectrum of the kinetic operator written as a series, and another handling of the latter would be required. In \cite{BBC}, suppression of the second-derivative term in the d'Alembertian relative to the Hubble friction is assumed, and $\rme^\B\approx \rme^{-3H_0 \p_t}$. This approximation is valid only by virtue of the special kind of solutions considered, while it is not in general.


\subsection{Problems and open issues}

The main virtue of the integral representation of nonlocal fields is to show that nonlocal theories are tractable in a quantitative way. However, the analytic and numerical inspections of the dynamical equations for the chosen example ($a= t^{H_0}$) are still incomplete and the fact that one or both equations of motion are not exactly solved may be subject to different interpretations.
\begin{enumerate}
\item[(i)] Extra (decoupled) particle fields might be invoked and their evolution equations integrated into the system, as for level corrections to the CSFT tachyon potential. However, such eventuality would not escape the problem of finding single-field nonlocal solutions given a local system as initial condition. 
\item[(ii)] The failure of the equations of motion might indicate that the local solution cannot be an initial condition for the nonlocal system. Other initial conditions might be concocted but this case would undermine one of the basis of the proposed method; that is, the possibility to attack nonlocality thanks to the systematic knowledge of the physics at $r=0$. 
\item[(iii)] Third, we have not considered modifications of the minimal system with ``evanescent'' terms of the form $h(\phi)-h(\tphi)$ or $r^n h(\phi,\tphi)$, $n>0$. These can be tuned so that to adjust the asymptotic behaviour at early times and may provide a necessary ingredient for the solution of the dynamical equations.
\item[(iv)] A fourth possibility is that the nonlocal cosmological system defined under the substitution $U(\phi)\to U(\tphi)$ in the local equations \emph{has no (analytic) solution except the local one}. We do not know whether this ``no-go'' conjecture is a general characteristic of nonlocal FRW systems or only of those with power-law scale factor. 
Counter-examples on a noncosmological (in particular, Minkowski) background are already known \cite{FGN,roll}.
\end{enumerate}

The space ${\cal S}_{\rm de}$ of functions obeying the diffusion equation includes states found via the direct computation method in the cases where the series is finite (i.e., for a particular choice of the parameters); we denote this subspace as ${\cal S}_*$. Therefore: {\bf (A)} either the space ${\cal S}_{\rm phys}$ of physical states (that is, those satisfying the equations of motion) is a subset of the direct approach for finite series (${\cal S}_{\rm phys}\subset{\cal S}_*$), or {\bf (B)} the solutions obey the diffusion equation but are infinite series (${\cal S}_{\rm phys}\subset{\cal S}_{\rm de}\setminus{\cal S}_*$), or {\bf (C)} they lay outside ${\cal S}_{\rm de}$.
While (C) is beyond the scope of this paper, we did not yet found any conclusive result about (A) and (B). In fact, the particular examples we considered (power-law universes) seem not to have any nonlocal nontrivial solution, but there is no evidence that such a negative situation is a general feature of nonlocality on FRW. On the other hand, solutions of the string equations of motion do belong to case (B) \cite{FGN,roll}.

There are several open problems which should be part of the future research agenda. Among them:
\begin{enumerate}
\item[(i)] Models with other choices for the Hubble parameter and the local potential should be considered. Superaccelerating examples [Eq.~\Eq{aexp}] and the case of the CSFT tachyon are of  particular relevance \cite{cuta3}.
\item[(ii)] The dynamical system can be studied also within the Hamilton--Jacobi formalism, where the coordinate $t$ is traded for the scalar field as a measure of time. In particular, the Friedmann equation \Eq{FRWth} can be simplified into \cite{cutac}
\be\label{hj2}
\left[1-\frac{\epsilon(2-\theta)}{6}\right]\frac{H^{2-\theta}}{\beta_\theta^2}+\cO_1-\tilde V=0,
\ee
where the dependence on $\cO_2$ and $\dot\phi$ has disappeared and $\phi$ is regarded as the natural clock of the system. A preliminary check shows that our numerical results are not modified in this framework.
\item[(iii)] The classical (i.e., phase-space) stability of the nonlocal solutions is to be assessed.
\item[(iv)] Cosmological linear perturbations would be important to describe inflationary observables as well as semiclassical stability of nonlocal models. Even if the Minkowskian theory is ghost-free, this is not guaranteed on a curved background. However, we expect that small-scale perturbations are unchanged in nonlocal scenarios. Consider the perturbation equation Eq.~(4.41) of \cite{cutac} for a test field (no metric backreaction). When the comoving spatial momentum of the perturbation is large, $|{\bf k}|\to \infty$, the effective mass term is subdominat and the harmonic oscillator equation of standard inflation is recovered.
\item[(v)] We have not said much about the singularity problem. Are there nonlocal dynamics which are smooth at the big bang?
\item[(vi)] Since the scalar field behaves as a perfect fluid, Eq.~\Eq{FRWth} relies upon the same assumptions advocated in the local braneworld. The first is that there is a confinement mechanism such that matter lives on the brane only, while gravitons are free to propagate in the empty bulk. This is guaranteed as long as the brane energy density $\rho$ is small compared to the effective gravitational 5D mass coupling. Then the energy-momentum tensor is covariantly conserved and the continuity equation, i.e., the scalar equation of motion, has no source term (corresponding to the 05-component of the bulk tensor \cite{VDBL,VDMP,LSR,KKTTZ,LMS,BiBC,ApT,tet04}). This does not imply that the contribution of the Weyl tensor is neglected (the second assumption): given a standard continuity equation on the brane (empty bulk), the Friedmann equation still can get an extra dark-radiation term. In the local braneworld, bulk physics mainly affects the small-scale or late-time (i.e. post-inflationary) cosmological structure and can be consistently neglected during the early history of the universe. For local inflation, this is a highly nontrivial result which has been confirmed with several methods both analytically and numerically (\cite{KMRWH} is among the most recent instances). Intuitively, the Randall--Sundrum dark-radiation term, which is the simplest contribution of the Weyl tensor, scales as $a^{-4}$ and is damped during the expansion of the universe. 

If the gravitational sector is local, these results do not depend on the matter content as long as inflation is enforced. If it is not (a possibility indicated by our findings), only an investigation of a concrete model would set the issue. However, a dark-radiation term in the Friedmann equation would not affect the shape and parameters of the nonlocal solutions, because these are determined by the scalar equation. In principle it can modify the early-time behaviour, but in that region the functions we found cease to be solutions of the Friedmann equation. We checked this even in the presence of a simple Weyl contribution $\propto a^{-4}= t^{-4H_0}$. In this case, moreover, the accuracy of the solution is considerably improved at intermediate times and the confidence value $t_*$ (Sec.~\ref{solutionBW}) is decreased.

The validity of the simple configuration of a brane with nonlocal 4D matter in an empty 5D bulk depends on the details of the model and cannot be decided by heuristic considerations. One of the problems is the implementation of nonlocality also in the gravitational action (see below).
\item[(vii)] It would be interesting to quantify the relation between the heat equation approach and the $1+1$ Hamiltonian formalism \cite{LV}. The main difference is that the diffusion equation is second order in time derivatives and first order in $r$-derivatives, while the auxiliary variable in the $1+1$ formalism acts as a time translation (first-order equation). So it should be understood how to incorporate the diffusion equation as a covariant constraint in the Lagrangian.
\item[(viii)] We already mentioned the necessity to extend our construction to instantonic solutions. This may be related to another aspect constituting new territory of research in cosmology: boundary string field theory (BSFT), for which solutions now have been constructed which are in deep relation with those of CSFT \cite{roll}. One can extend the calculation of the BSFT energy-momentum tensor of \cite{lar02} to an FRW background and study the resulting Friedmann equation.
\item[(ix)] It should be underlined that to put a curved metric into the original effective action of CSFT is not as straightforward as it may seem. A gravitational background is mediated by the 2-tensor modes of the closed-string perturbative spectrum (defined on the stable vacuum), while in the initial configuration only open-string modes (including the tachyon) live on the unstable brane. The direct substitution $\eta_{\mu\nu}\to g_{\mu\nu}$ into the open-string effective action is certainly an intuitive step helpful in the description of cosmological scenarios, but further meditation on this issue according to the rules of string theory would be of benefit.
\item[(x)] For consistency, one should consider the effect of nonlocal interactions also in the gravitational sector and check how the dynamics discussed in this paper is modified. The nonperturbative string formulation of gravity is still unknown, and only phenomenological \emph{Ans\"{a}tze} have been proposed so far \cite{BMS,kho06}. One possibility would be to adopt an educated guess about the form of the effective gravitational action and extend the heat equation formalism to its degrees of freedom. The task seems lengthy but at hand; of course, its results would be sensitive of the choice of the action. Another avenue to explore might be to truncate the nonlocal scalar part to finite order in $\alpha'$ and couple it to gravity at the same order. However, it seems unlikely that a perturbative approximation would shed new insights in the problem. The reason is that, due to higher-order differential operators, ghosts and other instabilities would appear in all sectors. These instabilities have been studied at second order in $\alpha'$ (Gauss--Bonnet modulus gravity) for an FRW background in \cite{CDD} (see also \cite{KM1,KM2,LN,DH}). One way to escape them might be to consider very realistic models with particular potentials and matter content. All of these ingredients should be dictated by string theory and under careful control throughout the evolution of the universe. The price to pay for abandoning the nonperturbative regime would be an extremely complicated model whose features would be accessible only numerically.
\end{enumerate}


\begin{acknowledgments}

G.C. thanks T. Biswas, L. Boyle, R. Brandenberger, J. Cline, L. Kofman, A. Liddle, A. Lukas, T. Naskar, S. Panda, M. Sami, and P. Singh for useful discussions. G.N. thanks the Department of Physics at Trento for kind hospitality. G.C. is supported by a Marie Curie Intra-European Fellowship under Contract No. MEIF-CT-2006-024523.

\end{acknowledgments}


\appendix*


\section{Hypergeometric functions}

We collect some properties of hypergeometric functions, widely used in Secs.~\ref{1tbw} and \ref{1t4d}. In the following and except in Eq.~\Eq{dif2}, $\a$ and $\b$ are arbitrary constants which have nothing to do with the coefficients in the nonlocal equations of motion.

The equation
\be\label{kumeq}
z\frac{\rmd^2 u}{\rmd z^2}+\left(\b-z\right) \frac{\rmd u}{\rmd z}-\a u=0\,,
\ee
admits two solutions (\cite{GR}, formul\ae\ 9.216.1-3), which are Kummer's confluent hypergeometric functions of the first and second kind: $u_1=\Phi(\a;\,\b;\,z)$, $u_2=\Psi(\a;\,\b;\,z)$. $u_1$ (also called $_1F_1$ or $M$) is defined as
\be\label{kumm}
\Phi(\alpha;\,\beta;\,z)\equiv\sum_{k=0}^{+\infty} \frac{(\alpha)_k}{(\beta)_k } \frac{z^k}{k!}=\frac{\Gamma(\beta)}{\Gamma(\alpha)}\,\sum_{k=0}^{+\infty} \frac{\Gamma(\alpha+k)}{\Gamma(\beta+k)}\, \frac{z^k}{k!}\,,
\ee
where $(\a)_{k}=\Gamma(\a+k)/\Gamma(\a)$ is the Pochhammer symbol. Kummer's function of the second kind $\Psi$ (also called $U$) is:
\ba
\Psi(\a;\,\b;\,z)&=&\frac{\pi}{\sin \pi\b}\left[
\frac{\Phi(\alpha;\,\b;\,z)} {\Gamma(1+\alpha-\b)\Gamma(\b)}\right.\nonumber\\
&&\left.- z^{1-\b} 
\frac{\Phi(1+\alpha-\b;\,2-\b;\,z)}{\Gamma(\alpha) \Gamma(2-\b)}\right],\label{psikum}
\ea
which satisfies the identity
\be
\Psi(\a;\,\b;\,z)=z^{1-\b} \Psi(\a-\b+1;\,2-\b;\,z)\,.
\ee
To link this discussion with the physical system under study, we notice that the diffusion equation can be written as an operator acting on the two independent variables $z$ [Eq.~\Eq{zvar}] and $r$:
\be\label{dif2}
\frac{r}{\a}(\a\rmd_r-\B)=z\p_z^2+(1-\nu-z)\p_z+r\p_r\,,
\ee
which reproduces Eq.~\Eq{kumeq} for $u(r,z)=r^{-\a} \bar u (z)$. The solutions \Eq{good} ($\a=\nu-\theta/4$, $\b=1+\nu$) and \Eq{good2} ($\a=-\theta/4$, $\b=1-\nu$) are of this form.

The properties of Kummer's functions when $\a$ and/or $\b$ are integers are reported in the main body of the paper. Another relation of interest is the large $|z|$ expression
\be\label{largez}
\frac{\Phi(\alpha;\,\beta;\,z)}{\Gamma(\b)}\sim \frac{(-z)^{-\a}}{\Gamma(\b-\a)}+\frac{\rme^z z^{\alpha-\beta}}{\Gamma(\alpha)}\,.
\ee
To get this formula, we take the integral representation of $\Phi$
\ba
&&\Phi(\a;\,\b;\,z)\nonumber\\
&&\quad=\frac1{2\pi\rmi}\int_{-\rmi\infty}^{+\rmi\infty} \rmd s\, \frac{\Gamma(\a+s)}{\Gamma(\a)}\frac{\Gamma(\b)}{\Gamma(\b+s)}\Gamma(-s) (-z)^s\,,\nonumber\\
\ea
where the poles of $\Gamma(\a+s)$ are on the left of the integration contour and those of $\Gamma(-s)$ are on its right.
If $|z|<1$, the contour is closed to the right and one gets Eq.~\Eq{kumm}, which is then analytically continued to all $z$. If $|z|>1$, one can close the contour to the left, and compute the integral by taking the residues at the poles of $\Gamma(\a+s)$: 
\[
{\rm Res}[\Gamma(s+\a),\,s=-\a-k]=\frac{(-1)^k}{k!}\,,\qquad k\in\mathbb{N}\,.
\]
The result is
\ba
\Phi(\a;\,\b;\,z)& \stackrel{z\to-\infty}{\sim}& (-z)^{-\a} \sum_{k=0}^{M}\frac{\Gamma(\a+k)}{\Gamma(\a)}\nonumber\\
&&\quad\times\frac{\Gamma(\b)}{\Gamma(\b-\a-k)}\frac{z^{-k}}{k!}\,,\label{dive}
\ea
which would not converge if $M=+\infty$. In practice, one truncates the series up to some finite order $M$. The first contribution in Eq.~\Eq{largez} is the leading term $k=0$. The second (subleading when $z<0$, dominant otherwise) can be inferred via Kummer's transformation
\be\label{kumtra}
\Phi(\alpha;\,\beta;\,z)=\rme^{z}\Phi(\beta-\alpha;\,\beta;\,-z)\,,
\ee
giving
\ba
\Phi(\a;\,\b;\,z) &\stackrel{z\to+\infty}{\sim}& \rme^{z}z^{\a-\b} \sum_{k=0}^{M}\frac{\Gamma(\b-\a+k)}{\Gamma(\b-\a)}\nonumber\\
&&\quad\times\frac{\Gamma(\b)}{\Gamma(\a-k)}\frac{(-z)^{-k}}{k!}\,,\label{dive2}
\ea
where we have truncated the series to order $M$.

Combining these formul\ae\ in Eq.~\Eq{psikum} and using the identity
\[
\frac{\Gamma(1-\a+k)}{\Gamma(1-\a)}=(-1)^k \frac{\Gamma(\a)}{\Gamma(\a-k)}\,,
\]
one also gets
\be\label{psidive}
\Psi(\a;\,\b;\,z)\ \stackrel{z\to\pm\infty}{\sim}\ z^{-\a} \sum_{k=0}^{M}(\a)_k(1+\a-\b)_k\frac{(-z)^{-k}}{k!}\,.
\ee
The phases of Eqs.~\Eq{2infy-} and \Eq{2infy+} can be inferred as follows. One chooses the phase of $z$ near the branch cut of $\Psi$ and the position of the latter. Let them be $\rme^{\rmi\pi}$ and the negative real axis. From the above equation, $\psi_2\sim t^{\theta/2} z^{\theta/4} z^{-\theta/4}=t^{\theta/2}\rme^{n\rmi\pi \theta/4}$. When $z>0$, $n=0$, while when $z<0$ the phases of $z^{\theta/4}$ and $z^{-\theta/4}$ have to be taken above and below the branch cut, so that $n=2$.

Formul\ae\ 13.4.10 and 13.4.23 of \cite{AS} are helpful to write the equations of motion in a simple form:
\ba
(\a+z\p_z)\Phi(\a;\,\b;\,z) &=& \a\Phi(\a+1;\,\b;\,z)\,,\label{fophi}\\
(\a+z\p_z)\Psi(\a;\,\b;\,z) &=& \a(1+\a-\b)\Psi(\a+1;\,\b;\,z)\,.\nonumber\\\label{fopsi}
\ea
The derivative of $\Phi$ with respect to its first argument is used to get the solutions in a four-dimensional power-law universe:
\be\label{deal}
\p_\a \Phi(\a;\,\b;\,z)=\frac{\Gamma(\b)}{\Gamma(\a)}\sum_{k=1}^{+\infty} [\psi(\a+k)-\psi(\a)]\,\frac{\Gamma(\a+k)}{\Gamma(\b+k)}\,\frac{z^k}{k!}\,,
\ee
where $\psi$ is the digamma function and
\be\label{useful}
\psi(\nu+k)-\psi(\nu)=\sum_{n=1}^k \frac{1}{\nu+n-1}\,.
\ee
When $\nu=1$, $H_k\equiv \psi(1+k)-\psi(1)=1+1/2+1/3+\cdots+1/k$ is called the \emph{harmonic number}.

Another function used in this paper is the generalized hypergeometric function
\be\label{2F2}
_2F_2(\a,\,\b;\,\tau,\,\delta;z)\equiv \sum_{k=0}^{+\infty} \frac{(\a)_k (\b)_k}{(\tau)_k (\delta)_k} \frac{z^k}{k!}\,.
\ee


\end{document}